\documentclass{article}

\usepackage{arxiv}

\usepackage{latexsym}
\usepackage{graphicx}
\usepackage{multicol,multirow}
\usepackage{amsmath,bm,amssymb,amsfonts}
\usepackage{mathrsfs}
\usepackage{amsthm}
\usepackage{algorithmic}
\usepackage{algorithm}
\usepackage{apacite}
\usepackage{rotating}
\usepackage{appendix}
\usepackage[authoryear]{natbib}
\usepackage{ifpdf}
\usepackage[T1]{fontenc}
\usepackage{times}
\usepackage{sourcesanspro}
\usepackage{newtxmath}
\usepackage{textcomp}%
\usepackage{xcolor}%
\usepackage{hyperref}
\usepackage{caption}
\usepackage{subcaption}
\usepackage{mathtools}
\usepackage{tabularx}

\usepackage[utf8]{inputenc} 
\usepackage[T1]{fontenc}    
\usepackage{url}            
\usepackage{booktabs}       
\usepackage{nicefrac}       
\usepackage{microtype}      
\usepackage{lipsum}		

\defcitealias{AIS-IMO:2003}{IMO, 2003}

\title{A Multivariate Space-Time Dynamic Model for Characterizing the Atmospheric Impacts Following the Mt. Pinatubo Eruption}

\author{Rob Garrett\\
	University of Illinois, Urbana-Champaign\\
	Champaign, Illinois\\
	Sandia National Laboratories\\
	Albuquerque, NM \\  \\
	\And
	\href{https://orcid.org/0000-0002-5239-5185}{ \includegraphics[scale=0.06]{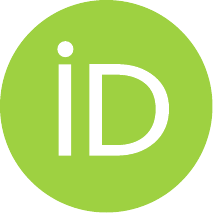}\hspace{1mm}Lyndsay Shand} 
	\thanks{corresponding author, lshand@sandia.gov} \\
	Sandia National Laboratories\\
	Albuquerque, NM \\	\\
	University of Illinois, Urbana-Champaign\\
	Champaign, Illinois\\
	\And
	\href{https://orcid.org/0000-0003-1651-1430}{ \includegraphics[scale=0.06]{orcid.pdf}\hspace{1mm}J. Gabriel Huerta}\\
	Sandia National Laboratories\\
	Albuquerque, NM \\ \\
	University of New Mexico\\
	Albuquerque, NM\
}

\renewcommand{\shorttitle}{\textit{arXiv} Template}

\begin{document}
\maketitle

\begin{abstract}
The June 1991 Mt. Pinatubo eruption resulted in a massive increase of sulfate aerosols in the atmosphere, absorbing radiation and leading to global changes in surface and stratospheric temperatures. 
A volcanic eruption of this magnitude serves as a natural analog for stratospheric aerosol injection, a proposed solar radiation modification method to combat a warming climate.
The impacts of such an event are multifaceted and region-specific. Our goal is to characterize the multivariate and dynamic nature of the atmospheric impacts following the Mt. Pinatubo eruption. 
We developed a multivariate space-time dynamic linear model to understand the full extent of the spatially- and temporally-varying impacts.  
Specifically, spatial variation is modeled using a flexible set of basis functions for which the basis coefficients are allowed to vary in time through a vector autoregressive (VAR) structure. 
This novel model is caste in a Dynamic Linear Model (DLM) framework and estimated via a customized MCMC approach. 
We demonstrate how the model quantifies the relationships between key atmospheric parameters prior to and following the Mt. Pinatubo eruption with reanalysis data from MERRA-2 and highlight when such model is advantageous over univariate models.
\end{abstract}


\section{Introduction}\label{sec:intro}

\begin{figure}[htbp]
\centering
\includegraphics[width=\linewidth]{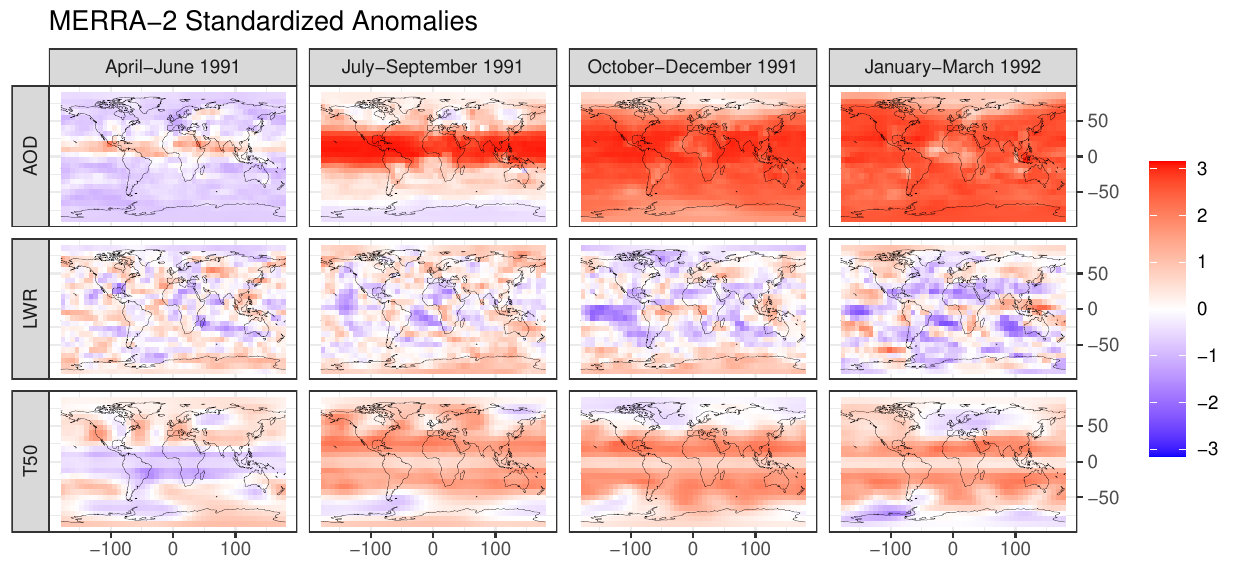}
\caption{Quarterly mean summaries of aerosol optical depth (AOD), longwave radiation (LWR), and stratospheric temperature at the 50mb pressure level (T50) for one year following the Mt. Pinatubo Eruption. Each variable is obtained from the MERRA-2 Reanalysis dataset and converted to standardized anomalies as detailed in Section \ref{sec:data} before taking means over the indicated months. Color fill is used to denote the standardized anomaly at each location.}
\label{fig:merra2_data}
\end{figure}

Long-term shifts in weather patterns, such as the significant warming of the earth’s surface and an increase in extreme events such as droughts and wildfires have become harder to ignore. Recently, many methods have been proposed to combat global warming, including mitigation (e.g. cutting emissions) and climate intervention (e.g. solar radiation modification, SRM). 
A promising form of SRM is stratospheric aerosol injection (SAI). SAI proposes to inject large amounts of aerosols into the stratosphere with the goal of cooling surface temperatures globally via changing the earth's radiation balance. For an overview of benefits and risks of SAI, see \cite{robock2009}.
Although, no large-scale SAI deployment has been demonstrated, there is a need to understand and potentially anticipate the numerous, region-specific impacts such an event would have. 
Natural events, such as volcanic eruptions, can mimic such an event under the right circumstances \citep{robock2000, rasch2008}.
The June 1991 Mt. Pinatubo eruption, for example, resulted in a massive increase of sulfate aerosols in the atmosphere, absorbing radiation and thus warming the stratosphere and cooling the earth's surface nearly 2 years later \citep{mccormick1995, boretti2024}.
The motivation of this work is to develop a statistical method that can characterize the multivariate and dynamic nature of the global impacts following the Mt Pinatubo eruption. Understanding the spatial and temporal characteristics of multivariate relationships between key atmospheric parameters affected by the eruption will be important to anticipating near to far-term impacts. Figure \ref{fig:merra2_data}, for example, shows the stark changes in the standardized anomalies in three variables: aerosol optical depth (AOD), outgoing longwave radiation, and stratospheric temperatures following the eruption. There is a known physical dependency between these variables which is most apparent in the similar spatial patterns between AOD and stratospheric temperatures.

There is a rich literature of spatial methods that are scalable to the large scale environmental data inherent to this problem.
The low rank approach known as Fixed Rank Kriging (FRK) \citep[e.g. ][]{cressie2008, sainsbury2024} is a popular and easily implementable choice. Approaches such as Lattice Kriging by \cite{nychka2015}, Multi-Resolution Approximations by \cite{katzfuss2020}, and those summarized in \cite{cressie2022}, which include FRK, rely on basis functions as a spatial reduction tool and to induce sparsity in the covariance or precision matrix. The Nearest Neighbor Gaussian Process \citep{datta2016, finley2018} leverages the nearest neighbor likelihood evaluation demonstrated by \cite{vecchia1988} to estimate a latent Gaussian process. More recently, \cite{katzfuss2021} generalizes the work of  \cite{vecchia1988} to enable Gaussian Processes to be approximated more accurately. A comprehensive review of many of these methods can be found in \cite{heaton2019}. 
Further, there are a number of nonparametric approaches that have been proposed such as the scalable Bayesian Transport Map proposed by \cite{katzfuss2023}. A number of deep learning algorithms have also shown promise when prediction is the goal \citep{wikle_deeplearning2023}.

Some of these methods have been extended to model space-time data, typically in a state-space and/or a Bayesian Hierarchical framework \citep[see][]{banerjee2014}, \cite{stroud2001}, \cite{huerta2004} and \cite{gelfand2005} were some of the first to consider the dynamic modeling framework to represent space-time processes. Further, \cite{finley2012bayesian} proposed a Bayesian dynamic model for space-time data using predictive processes as in \cite{finley2009}. \cite{katzfuss2012bayesian} proposed a large-scale spatio temporal smoothing method via a Bayesian multiresolution basis representation. Additionally, \cite{jurek2023} presents a fast approximation to the forward filter backward sampler (FFBS) commonly used in the Bayesian sampling scheme for space-time models. For a thorough review of the literature on hierarchical dynamic spatio-temporal models, see \cite{cressie2011_ch7}. Moreover \cite{wikle_deeplearning2023} also demonstrate the usefulness of statistical deep learning methods for predicting space-time processes such as echo state networks \citep{mcdermott2017, mcdermott2019_esn} and recurrent neural networks \citep{mcdermott2019_rnn} to name a few.

Fewer available methods can handle multivariate space-time processes, likely due to the high-dimensional nature of multivariate space-time processes and the tendency for such model specifications to become over parameterized. 
\cite{gelfand2005} proposed a dynamic framework to model the multivariate process of precipitation and temperature data across space and time. Although this approach is capable of capturing multivariate dependencies, these dependencies are not allowed to vary across space. 
\cite{bradley2015} proposed a multivariate space-time mixed effects model for high-dimensional areal data while \cite{kleiber2019} proposed a multiresolution approach to ensure scalability and capture multivariate correlations within the spatial covariance matrix.
There have been some multivariate spatial methods proposed that are very challenging to extend to space-time due to their high computational demand such as those reviewed in \cite{gelfand2021}, e.g. cokriging methods and models of coregionalization. 
In addition, \cite{banerjee2008} proposed a fully Bayesian spatial multivariate predictive process and suggest its scalability to space-time data.
For a more computationally efficient method, deep learning approaches would also be appropriate to model such high dimensional multivariate space-time processes but would not allow for interpretability of the multivariate relationships, a key interest for our application.

The impact of stratospheric aerosols on the Earth's radiative balance and thus on temperature is well documented but not well quantified due both the difficulty in studying atmospheric processes at this scale and high complexity of climate dynamics.
Due to our specific interest in understanding dynamic multivariate relationships in the atmosphere and the limited inventory of available multivariate space-time methods, we propose a novel Multivariate Space-Time Dynamic Model (MV-STDM) to capture the basic dependencies between AOD, outgoing longwave radiation, and stratospheric temperatures impacted by the Mt. Pinatubo eruption. 
Our MV-STDM incorporates a spatial basis function representation for scalability due to the high data dimension inherent in multivariate spatiotemporal fields data.
The main novelties of our work are the multivariate framework, which is motivated by the unique application to impacts following the Mt. Pinatubo eruption, and the estimation and interpretability of a spatially-varying transition matrix that captures temporal dependence in the basis function coefficients. Specifically, we introduce significant computational benefits through our intentional specification of the autoregressive coefficient matrix and clever vectorization within the estimation algorithms. Interpretability, and especially parameter identifiability, is challenging in most multivariate approaches because multivariate models are easily overspecified. We mitigate the risk of unidentifiability by parameterizing the model such that the temporal and spatial correlation are distinctly captured by separate components.
We estimate our model within a Bayesian hierarchical framework with an efficient Markov Chain Monte Carlo (MCMC) sampling algorithm. 
By fitting our model to observational data over a time period before and after the Mt. Pinatubo eruption, we find evidence of a change spatial trends in the transition matrix for many of the atmospheric processes.
We also find that our MV-STDM offers advantages over simpler univariate spatiotemporal models for predicting changes in stratospheric temperatures following the eruption.
More detailed results are given in Section \ref{sec:application}.

The remainder of this paper is organized as follows. Section \ref{sec:data} motivates our multivariate model by presenting the atmospheric parameters of interest. Section \ref{sec:model} presents details on our model formulation and describes the MCMC sampling procedure used for estimation. Section \ref{sec:simulation} demonstrates the robustness of our model under a simulated data scenario, and Section \ref{sec:application} presents the application to the motivating dataset.
Lastly, Section \ref{sec:discussion} provides a discussion of our findings as well as possible extensions and limitations of our work.

\section{MERRA-2 Data}\label{sec:data}

We study three different atmospheric variables that had notable observed changes following the 1991 Mt. Pinatubo eruption. These include aerosol optical depth (AOD), upwelling longwave radiation (LWR), and stratospheric temperature measured at 50mb (T50). This collection of variables represents an important interaction between atmospheric processes following the eruption, where the influx of stratospheric aerosols absorbed a greater than usual amount of outgoing longwave (infrared) radiation, leading to global stratospheric warming which reached up to $3^o$C near the equator \citep{angell1997}. We obtain each variable from the Modern-Era Retrospective analysis for Research and Applications, Version 2 (MERRA-2), which is a global reanalysis dataset \citep{gelaro2017}. 
For AOD, we obtain the total aerosol extinction aerosol optical thickness variable at a 550 nm wavelength, or the product labeled TOTEXTTAU. AOD is a unitless measurement. For LWR, we obtain the upwelling longwave flux at top of atmosphere variable, measured in $\text{W}/\text{m}^2$, or LWTUP. Finally, for stratospheric temperature, we obtain the air temperature at a pressure level of 50 mb, labeled as T50 and measured in Celsius. All three variables we obtained at a monthly temporal frequency and on a $1^{\circ}\times 1^{\circ}$ regular latitude-longitude grid ($360 \times 180$ spatial grid).

We apply a few preprocessing steps to the data. First, the spatial fields for each variable and time point are regridded to a $24 \times 48$ spatial grid through spatial averaging. Because our approach is focused on understanding multivariate relationships rather than providing fine-scale spatial predictions, this lower resolution helps to keep the dimension of our data reasonable while still capturing long range spatial trends with a reasonable number of spatial basis functions. Second, we subset a twelve year time period of 1984-1995, centered around the 1991 eruption, to have a sufficient number of time points with which to estimate our model parameters. This study time period was also chosen to avoid including a prior large volcanic eruption by El Chich{\'o}n in 1982. 
Lastly, to isolate the effects of the eruption from typical spatial or seasonal trends, we convert the observations for each variable into standardized anomalies. This first requires computing the the monthly mean and monthly standard deviation at each location over the 1984-1995 time period.
Standardized anomalies are then calculated for each observation and time by subtracting the monthly mean (aka climatologies) and dividing by the monthly standard deviation at that location. 
Figure \ref{fig:merra2_data}, referenced in the introduction, shows these anomalies over four 3-month periods before and after the eruption.

\section{Method}\label{sec:model}

\subsection{Space-Time Framework}

Dynamic linear models (DLMs) are a general class of models that provide a state space representation for time series as presented in \cite{petris2009} and \cite{West-Harrison97}. Recent adaptations of DLMs for spatiotemporal data can be found in, for example, \cite{rougier2023}, \cite{Wikle2019},\cite{datta2016}, \cite{banerjee2014}, \cite{finley2009}, \cite{finley2012bayesian}.
 Our modeling approach is motivated by \cite{rougier2023} which discusses the Kalman Filter as a ``workforce'' for dynamic modeling of complex environmental processes. 
We begin by introducing our DLM framework for spatiotemporal data, then extend this model to accommodate multivariate spatiotemporal fields. 

We consider the problem of modeling spatiotemporal data observed at $N$ locations over $T$ time periods. We denote the observations from the spatial field at time $t$ as the $N\times 1$ vector $\bm y_t$. Following the typical DLM structure, we begin by specifying an \textit{observation equation}, which models each $\bm y_t$ using a spatial basis function representation, and an \textit{evolution equation} for the basis coefficients, which captures spatial and temporal autocorrelation through the use of a transition matrix and spatially-correlated error structure. 

Our DLM framework is then written as
\begin{align}
\label{eq:dlm_uni}
 {\bm{y}}_t &= {\bm\Phi}{\bm{\alpha}}_t+{\bm{\epsilon}}_t, \quad {\bm{\epsilon}}_t \overset{\mathrm{ind.}}{\sim} N(\bm{0}_N,\sigma_t^2\bm I_N),  \\
 {\bm{\alpha}}_t &= A {\bm{\alpha}}_{t-1} + {\bm{\eta}}_t, \quad {\bm{\eta}}_t \overset{\mathrm{i.i.d.}}{\sim} N(\bm{0}_K,\tau^2(\bm B' \bm B)^{-1}), 
\end{align} 
where $\bm I_N$ is the $N\times N$ identity matrix, $\bm 0_N$ denotes the $N\times 1$ vector filled with 0, and $'$ is the matrix transpose operation. In the observation equation, $\bm \Phi$ is the $N\times K$ matrix containing $K$ Radial Basis Functions (RBFs), $\bm \alpha_t$ is the $K\times 1$ state vector of time-varying coefficients, and $\bm \epsilon_t$ is the $N\times 1$ vector of measurement errors at each location with variance parameterized by $\sigma^2_t$. In the evolution equation, $A$ is the $K\times K$ transition matrix which captures the transitions between states. 

We further constrain $A$ to be a diagonal matrix.  
Lastly, $\bm \eta_t$ is the $K\times 1$ vector of spatially-correlated innovation errors with covariance parameterized by a pre-specified $K\times K$ spatial autoregression (SAR) matrix $\bm B$ and a scale parameter $\tau^2$.

Both $\bm \Phi$ and $\bm B$ follow the specification introduced in \cite{nychka2015} for rectangular spatial geometries and later adapted for spherical spatial geometries in the `LatticeKrig' R package \citep{nychka2016lattice}. For our application, we use the spherical geometry representation to cover the global spatial domain of our data. Under this specification, our basis consists of $K$ Wendland RBFs each centered at a different location on an icosahedral grid over the Earth. We denote this grid as the `basis grid'. Following \cite{nychka2015}, each RBF is created using the following Wendland function
\begin{equation}
    \phi(d) = \begin{cases}
        (1-d)^6(35d^2+18d+3)/3 & 0\leq d \leq 1, \\
        0 & d > 1,
    \end{cases}
\end{equation}
where the distances $d$ are computed from each of the $N$ locations of the observations to the $K$ locations on the basis grid. The great-circle distance is used to accommodate the spherical structure, and a scaling factor is applied so that the range of each RBF extends approximately 2.5 times the distance between neighboring locations in the basis grid. 
Because these RBFs have finite range, the resulting basis matrix $\bm \Phi$ will be sparse.
For interpretability of our model parameters, particularly $A$, we consider only a single resolution of RBFs. Since the RBFs are associated with different regions, albeit overlapping, we capture spatial dependence within the innovation errors $\bm\eta_{t}$ by using the SAR matrix $\bm B$. For spherical geometries, $\bm B$ is a sparse matrix with each element varying depending on the adjacency structure of the basis grid. The $(i,j)^{th}$ element of $\bm B$ is equal to $1+\kappa^2$ if $i=j$, and equal to $-1/n_i$ if the $j$th location in the basis grid is adjacent to the $i$th location, and $0$ otherwise. $n_i$ denotes the number of locations adjacent to the $i$th location. Since the basis grid is created by recursively dividing the faces of an icosahedron, $n_i$ is either $5$ or $6$. The $\kappa$ parameter must be greater than 0 and its value determines the overall strength of the spatial correlation, with smaller values enforcing greater correlation between neighboring RBFs. See \cite{nychka2016lattice} for details and code implementation.

\subsection{Multivariate Extension}
\label{sec:method_multi}

We now extend our model to handle the multivariate case. 
For variable $i = 1,\ldots,M$, let $\bm y^{(i)}_t$ denote the $N \times 1$ vector of observations from the spatial field at time $t$. 
We jointly represent the observations $\bm y^{(i)}_t$, along with its corresponding measurement errors $\bm\epsilon_t^{(i)}$, state vector $\bm \alpha_t^{(i)}$, and innovation errors $\bm\eta_t^{(i)}$, using the following structures:
\begin{equation}
\label{eq:multi_vecs}
{\bm{Y}}_t= \left[\begin{array}{c}{\bm y}_t^{(1)}\\ \vdots \\ {\bm y}_t^{(M)}\end{array}\right],\qquad 
{\bm{E}}_t= \left[\begin{array}{c}{\bm\epsilon}_t^{(1)}\\ \vdots \\ {\bm\epsilon}_t^{(M)}\end{array}\right], \qquad
{\bm{\alpha}}_t= \left[\begin{array}{c}{\bm \alpha}_t^{(1)}\\ \vdots \\ {\bm \alpha}_t^{(M)}\end{array}\right], \qquad
{\bm{H}}_t= \left[\begin{array}{c}{\bm\eta}_t^{(1)}\\ \vdots \\ {\bm\eta}_t^{(M)}\end{array}\right].
\end{equation}
As a result of stacking the parameters for each variable, the observations and measurement errors, ${\bm{Y}}_t$ and ${\bm{E}}_t$, are now each of size $MN\times 1$, and the state vector and innovation errors, $\bm\alpha_t$ and $\bm H_t$, are now of size $MK\times 1$. We adjust the remaining model parameters to accommodate the increase in dimension size. The basis, measurement variance, and innovation variance matrices are expanded to match the increased model dimension as follows
\begin{equation}
\label{eq:multi_pars}
     \bm \Phi_M = \bm I_M \otimes \bm\Phi, \qquad \bm V_t = \text{diag}\left(\sigma^2_{1t},\ldots,\sigma^2_{Mt}\right) \otimes \bm I_n, \qquad \bm Q = \text{diag}\left(\tau^2_{1},\ldots,\tau^2_{M}\right)\otimes (\bm B'\bm B)^{-1},
\end{equation}
where $\otimes$ denotes the Kronecker product and $\text{diag}\left(\sigma^2_{1t},\ldots,\sigma^2_{Mt}\right)$, $\text{diag}\left(\tau^2_{1},\ldots,\tau^2_{M}\right)$ denote the $M \times M$ diagonal matrices with the elements $\sigma^2_{1t},\ldots,\sigma^2_{Mt}$ and $\tau^2_{1},\ldots,\tau^2_{M}$, respectively. 
We specify the same set of radial basis functions for each variable. Thus, $\bm\Phi_M$ is an $MN\times MK$ basis matrix which contains a copy of the original $\bm \Phi$ for each variable, $\bm V_t$ is the $MN\times MN$ diagonal measurement variance matrix which incorporates a different scale parameter $\sigma_{it}^2$ for each variable, and $\bm Q$ is the $MK\times MK$ innovation covariance matrix which keeps the same spatial autoregression matrix $\bm B$ for each variable, but allows the scale parameter $\tau^2_i$ to vary between variables. 

Note that neither $\bm V_t$ or $\bm Q$ incorporate multivariate dependencies. This is a purposeful choice to isolate multivariate relationships in the transition matrix, $\bm A$, thereby creating a more interpretable model. To match the enlarged dimension of the state vector, $\bm A$ is now an $MK \times MK$ matrix. We partition $\bm A$ into blocks to separate relationships between variables as follows
\begin{equation}
\label{eq:A}
{\bm A}= \left[\begin{array}{ccc}A_{11}& \dots &  A_{1M} \\ \vdots & \ddots & \vdots \\  A_{M1} & \dots &  A_{MM}\end{array}\right],
\end{equation}
where each block $A_{ij}$ is a $K\times K$ matrix that models the impact of $\bm\alpha^{(j)}_{t-1}$ on $\bm\alpha^{(i)}_t$. Since each element of $\bm\alpha^{(i)}_t$ corresponds to one RBF and thus one location in the basis grid, a fully dense specification of $A_{ij}$ would capture the dependence of variable $i$ on variable $j$ for every pair of locations in the basis grid.
For a dense specification of the $A_{ij}$ matrices, the resulting number of parameters in $\bm A$ would be $(MK)^2$, a potentially large number if even a modest resolution is chosen for $\bm \Phi$, thus presenting challenges for estimation and interpretation.
To maintain sparsity in our model specification while keeping important temporal dependencies in place, we constrain each block $A_{ij}$ for $i,j=1,\ldots,M$ as follows
\begin{align} 
\label{eq:A_diag}
A_{ij} = \text{diag}({\Tilde{ A}}_{ij}), \quad i, j \in \{1, \ldots, M\},
\end{align}
where each ${\tilde{A}}_{ij}$ is a $K\times 1$ vector. 
This constraint reduces the overall dimension of $\bm A$ to $M^2K$, and the dependence of variable $i$ on variable $j$ is now captured only between state vector coefficients which correspond to the same location in the basis grid. 

Finally, our multivariate space-time dynamic model (MV-STDM) incorporates the above parameterizations to provide a similar set of DLM equations to (\ref{eq:dlm_uni}) and is written as
\begin{align} 
\begin{split}\label{eq:dlm_full}
\bm Y_t & = \bm \Phi_M \bm \alpha_t + \bm E_t, \quad \bm E_t \overset{\mathrm{ind.}}{\sim} N(\bm 0_{NM}, \bm V_t), \\
\bm \alpha_t & = \bm A \bm \alpha_{t-1} + \bm H_t, \quad \bm H_t \overset{\mathrm{i.i.d.}}{\sim} N(\bm 0_{KM}, \bm Q),
\end{split}
\end{align}
where $\bm Y_t$, $\bm\alpha_t$, $\bm E_t$, and $\bm H_t$ follow the structure outlined in (\ref{eq:multi_vecs}), $\bm \Phi_M$, $\bm V_t$, and $\bm Q$ follow the specifications in (\ref{eq:multi_pars}), and $\bm A$ is subject to the sparsity constraint detailed in 
(\ref{eq:A_diag}). The evolution equation of the MV-STDM
can be seen as a first-order vector autoregressive, or VAR(1), structure that captures relationships in the evolution of the state vector for each variable.
Following typical assumptions for VAR models \cite{lutkepohl2005}, both the coefficients, $\bm A$, and the covariance, $\bm Q$, of the innovation process are fixed in time. The proposed model structure allows us to  partition the total variability in the data into random sources, such as measurement errors and spatial innovations, and deterministic sources, such as multivariate and temporal dependencies in the transition matrix. 

Interpretation of the blocks $A_{ij}$ of the transition matrix $\bm A$ is of particular interest to our application, allowing us to assess the strength and direction of multivariate temporal dependencies. We can project $A_{ij}$ onto the spatial domain of the observations via the product $\bm\Phi A_{ij}$. Due to our sparsity constraint, each $A_{ij}$ block has only $K$ terms corresponding to each RBF, so this allows us to visualize the relationship between $\bm\alpha^{(i)}_t$ and $\bm\alpha^{(j)}_{t-1}$ as a spatial map. To preserve the original scale of the coefficients in $ A_{ij}$, we divide each element of $\bm\Phi A_{ij}$ by the corresponding element of $\bm\Phi$.  This process is demonstrated in Figure \ref{fig:A_true_sim}, which depicts the blocks $A_{ij}$ both as points on the basis grid as well as projected onto the spatial domain. In Figures \ref{fig:APhi_application} and \ref{fig:APhi_application_ci}, 
we show only the $\bm \Phi A_{ij}$ representation as it provides a more clear depiction of the spatial trends. 

\subsection{MCMC Sampling}
\label{sec:mcmc}

Estimation of the MV-STDM in (\ref{eq:dlm_full}) is performed via a Markov Chain Monte Carlo (MCMC) method. 
We construct a Gibbs sampler to estimate the parameters which requires deriving full conditional distributions for each model parameter in (\ref{eq:dlm_full}).
In this section, we specify our prior distribution selections and respective posterior distributions. To faciliate the sampling, we specify conditionally-conjugate priors.
Our MV-STDM takes the standard form of a DLM, so conditional posterior sampling for the state vectors $\bm \alpha_0,\ldots,\bm \alpha_T$ 
can be performed using similar methodology to \cite{petris2009}, which relies on the Forward Filtering, Backward Sampling (FFBS) algorithm from \cite{carter1994} and \cite{fruwirth1994} to sample the state vector conditional on all other parameters. However, the transition matrix $\bm A$ has an additional sparsity constraint that is not common in the DLM literature. To account for this, we reformulate the evolution equation in (\ref{eq:dlm_full}) to sample the full conditional distribution of $\bm A$.

Following Bayesian VAR methods \citep{chan2020}, but making modifications for our specific model structure, we cast the transition equation as a Bayesian linear regression problem by reorganizing the structure of some parameters. 
First, we stack the $\bm\alpha_t$ and $\bm\eta_t$ vectors for every time point to obtain response and error vectors of size $TMK \times 1$. 
Then, we represent the nonzero terms of $\bm A$ as $\bm{\tilde{A}} = ({\tilde{A}}_{ij})$, the $MK \times M$ block matrix with blocks ${\tilde{A}}_{ij}$ for $i,j \in \{1,\ldots,M\}$. We take the column-wise vectorization of $\bm{\tilde{A}}$, denoted as $\text{vec}(\bm{\tilde{A}})$, to obtain a $M^2K \times 1$ vector of regression coefficients. Lastly, we build the design matrix of the linear regression model by specifying  $\bm X^\alpha_t = \bm I_M\otimes\begin{bmatrix}\text{diag}(\bm\alpha_t^{(1)}) & \ldots & \text{diag}(\bm\alpha_t^{(M)})\end{bmatrix}$ to reincorporate the diagonal structure of the blocks of $\bm A$. The kroncker product serves to increase the dimension of $\bm X^\alpha_t$ to $MK\times M^2K$ so that the number of columns matches the size of $\text{vec}(\bm{\tilde{A}})$. We can stack $\bm X^\alpha_0, \ldots, X^\alpha_{T-1}$ so that the number of rows matches the size of the stacked $\bm \alpha_t; t=1, \ldots T$.
Finally, the following regression model 
\begin{align} \label{eq:mvlr}
\begin{bmatrix}
    \bm\alpha_T \\
    \vdots \\
    \bm\alpha_1
\end{bmatrix} = 
\begin{bmatrix}
    \bm{X}^\alpha_{T-1} \\
    \vdots \\
    \bm{X}^\alpha_{0} \\
\end{bmatrix}
\text{vec}\left(\bm{\tilde{A}}\right) + 
\begin{bmatrix}
    \bm{\eta}_T \\
    \vdots \\
    \bm{\eta}_1
\end{bmatrix}, \quad \begin{bmatrix}
    \bm{\eta}_T \\
    \vdots \\
    \bm{\eta}_1
\end{bmatrix} \sim N\left(\bm 0_{TMK},\bm I_T \otimes \bm Q\right)
\end{align}
is equivalent to the evolution equation in  (\ref{eq:dlm_full}). 
For $\bm\alpha_0$, we specify a normal prior $\bm\alpha_0 \sim N(\bm m_0,\bm C_0)$ with $\bm m_0 = \bm 0_{MK}$ and $\bm C_0 = \bm I_{MK}$ following standard DLM assumptions in Petris et al. (2009).
\begin{align}
    {\tilde{A}}_{ij} &\overset{ind}{\sim} \begin{cases}
    N(\bm 1_K, \lambda \bm I_k), \quad i=j,\\
    N(\bm 0_K, \lambda \bm I_k), \quad i\neq j,
\end{cases}
\end{align}

and the sampling of the full conditional posterior of  $\bm A$ can be completed via Bayesian linear regression method as shown in \cite{chan2020} for the general Minnesota prior. The assigned prior distribution for $\bm\alpha_0$ follows common practice for the state vector at time $0$ in DLMs. The prior distribution for ${\tilde{A}}_{ij}$ is a special case of the Minnesota prior, which is popular for Bayesian VAR models \citep{lutkepohl2005}.
In the case that $i=j$, the prior mean is $1$, which supports our belief that there will be strong positive autocorrelation for each variable over time. In the case that $i\neq j$, the prior mean is $0$, which allows for either positive or negative cross-correlations over time. In practice, small values for the the prior variance, $\lambda$, serve to shrink the coefficients of ${\tilde{A}}_{ij}$ closer to their prior mean.

The prior distributions for the variance terms are specified as follows:
\begin{equation}
    \sigma^2_{it} \overset{ind}{\sim} IG(a_\sigma,b_\sigma), \qquad \tau^2_i \overset{ind}{\sim} IG(a_\tau,b_\tau),
\end{equation}
where $IG$ denotes an Inverse-Gamma distribution. Here, the prior distributions for $\sigma_{it}^2$ and $\tau^2_i$ allow for Normal-Inverse Gamma conjugacy to obtain the corresponding full conditional posterior distributions which will also be Inverse-Gamma. 

Full technical details of the Gibbs sampling procedure are available in Appendix \ref{app:implementation}. We implemented this procedure in Rcpp \citep{eddelbuettel2011} using the `Armadillo' package for matrix operations \citep{eddelbuettel2014}. Sampling for the state vector $\bm\alpha_t$ is based on a modified version of the FFBS algorithm in the `dlm' R package \citep{petris2010}. Their code is not intended for a high dimensional state vector such as the one corresponding to our MV-STDM, so in our implementation of FFBS, modifications were made to account for the sparsity structure of the model parameters.  
The full implementation of our method is available at \url{https://github.com/garretrc/MV-STDM}.

\section{Simulation Study}\label{sec:simulation}

\subsection{Synthetic Dataset}
\label{sec:sim_data}

A simulation study was performed to test the effectiveness of our sampling algorithm and identifiability of our model parameters. 
We generate a synthetic data from our MV-STDM in (\ref{eq:dlm_full}) with $N=1,152$ locations, $T=144$ time points, and $M=3$ variables to mimic the dimension of the MERRA-2 reanalysis data described in Section \ref{sec:data}. For the basis functions $\bm\Phi$, we construct $K=42$ RBFs on an icosahedral grid.
Chosen for ease of computation, this is a lower basis resolution than we use for the main analysis in Section \ref{sec:application}.
For the innovation covariance $\bm Q$, we specify $\tau^2_i=5$ for all variables $i=1,2,3$, and generate the spatial autoregression matrix $\bm B$ using a value of $\kappa=2$.
To ensure the simulated data have sufficient spatial signal, we specify $\sigma^2_{it} = 2$ for all variables $i=1,2,3$ and all time points $t=1,\ldots,T$ so that the measurement errors, driven by $\bm V_t$, have a smaller impact on the process than the innovations, which are driven by $\bm Q$. 

\begin{figure}[h]
\centering
\includegraphics[width=\linewidth]{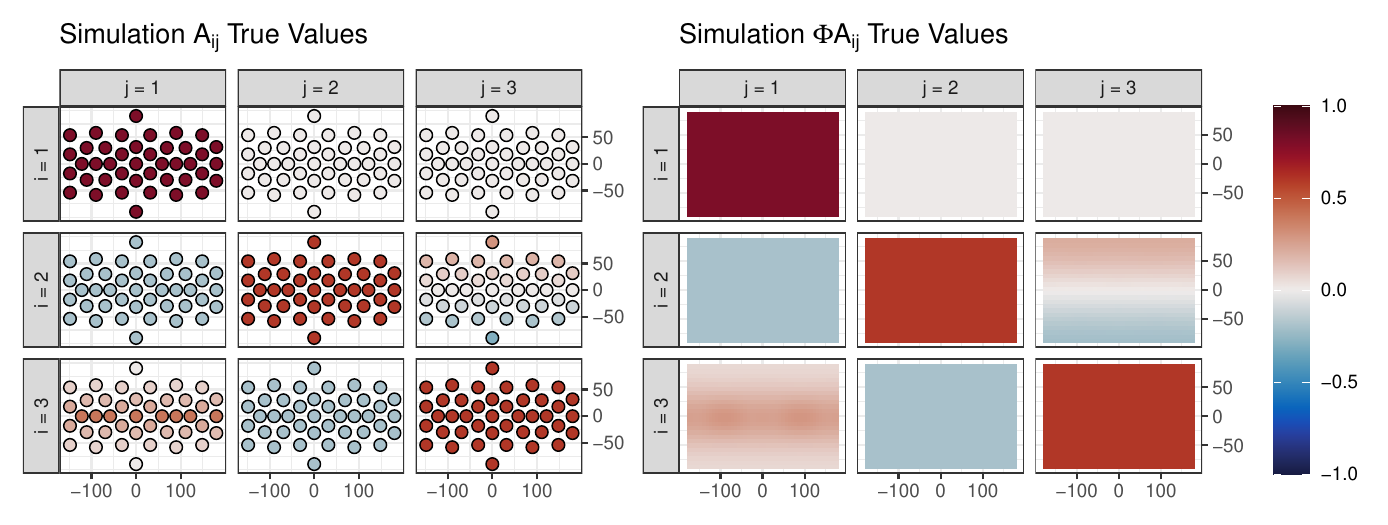}
\caption{
True values for the transition matrix, $\bm A$, used to generate our synthetic data example. The left panel shows each of the blocks $A_{ij}$ of $\bm A$, which represent the temporal dependency of variable $i$ on variable $j$ for each combination $i,j = 1,2,3$. These dependencies are allowed to vary in space across the 42 locations in the basis grid, represented as points with a color fill that corresponds to the parameter value. The right panel shows the same parameters projected onto the full spatial domain by taking the multiplication $\bm\Phi A_{ij}$ and adjusting the scale as detailed in Section \ref{sec:method_multi}. The color scale is shared for each panel.
}
\label{fig:A_true_sim}
\end{figure}

For the transition matrix, $\bm A$, we specify 
varying temporal correlation strengths for the blocks $A_{ij}$, each of which determines the temporal dependence of variable $i$ on variable $j$. Figure \ref{fig:A_true_sim} shows the chosen parameter values for $\bm A$.
The diagonal blocks $A_{ii}$ represent the autocorrelations or temporal dependencies of each variable on themselves. For these blocks, we specify strong positive values. Whereas, for the off-diagonal blocks $A_{ij}$, $i\neq j$, which represent the temporal cross dependencies of each variable on the other variables, we consider a variety of coefficients and spatial patterns. See Appendix \ref{app:simulation} for full details on how we constructed each block of $\bm A$.

Using the specified parameter values, we generate the state vectors $\bm \alpha_t$ from the model equations in (\ref{eq:dlm_full}). First, $\bm \alpha_0$ is initialized by taking a random draw from a standard normal distribution $N(\bm 0_{MK}, \bm 1_{MK})$. Each subsequent $\bm \alpha_t$ is generated recursively using the transition equation using the above specified values of $\bm A$ and $\bm Q$. The initial state vector $\bm\alpha_0$ is randomly drawn and does not exhibit the desired multivariate spatiotemporal dependencies. The subsequent state $\bm\alpha_1$ and future states are influenced by this arbitrary initialization. To mitigate this influence, we simulate the process until a total of 244 $\bm\alpha_t$ vectors are generated. The first 100 vectors are then discarded, yielding $T=144$ time points that more accurately reflect the desired multivariate spatiotemporal dynamics. For each $\bm \alpha_t$, the corresponding synthetic observations $\bm Y_t$ are generated directly from the measurement equation with the error terms $\bm\epsilon_t$ and $\bm\eta_t$ generated randomly using covariance matrices $\bm V_t$ and $\bm Q$, respectively.

\subsection{Simulation Results}

To test our MCMC sampler, we fitted our MV-STDM to the simulated $\bm Y_t$ data and compared the results to the known values of $\sigma^2_{it}$, $\tau^2_i$, and $\bm A$. 
We considered a prior with $\bm m_0 = \bm 0_{MK}$, $\bm C_0 = \bm I_{MK}$, $a_\sigma = b_\sigma = 1$, $a_\tau = b_\tau = 1$, and $\lambda = 1/4$. We produced a single chain of our MCMC of $2,500$ iterations. Burn in was diagnosed within $250$ iterations using trace plots, but we dropped the first $500$ to obtain a final set of $2,000$ posterior samples for each parameter.

\begin{table*}[h]
\caption{Posterior means and credible intervals for parameters estimated from the synthetic dataset. Columns correspond to each variable, denoted by indices $i=1,2,$ and $3$. Results are given for the innovation variance parameter, $\tau^2_i$, as well as the measurement error parameters, $\sigma^2_{it}$, at three time points, $t=1,72,144$. The lower and upper bounds of each credible interval are based on the 0.025 and 0.975 quantiles of the posterior samples, respectively.}
\label{tab1}
\centering
\begin{tabular*}{500pt}{cccccccc}
\toprule
& & \multicolumn{2}{c}{$i=1$} & \multicolumn{2}{c}{$i=2$} & \multicolumn{2}{c}{$i=3$} \\
\cmidrule{3-4} \cmidrule{5-6} \cmidrule{7-8}
& \textbf{True Value} & \multicolumn{1}{c}{\textbf{Post. Mean}} & \multicolumn{1}{c}{\textbf{95\% Post. CI}} & \multicolumn{1}{c}{\textbf{Post. Mean}} & \multicolumn{1}{c}{\textbf{95\% Post. CI}} & \multicolumn{1}{c}{\textbf{Post. Mean}} & \multicolumn{1}{c}{\textbf{95\% Post. CI}} \\
\midrule
$\tau_i^2$ & 5 & $4.89$ & $(4.67,5.12)$ & $4.85$ & $(4.63,5.07)$ & $5.04$ & $(4.82,5.27)$ \\ 
$\sigma^2_{i,1}$ & 2 & $2.03$ & $(1.87,2.21)$ & $1.97$ & $(1.81,2.14)$ & $2.00$ & $(1.84,2.18)$ \\
$\sigma^2_{i,72}$ & 2 & $2.00$ & $(1.84,2.17)$ & $1.97$ & $(1.81,2.15)$ & $2.11$ & $(1.94,2.30)$ \\
$\sigma^2_{i,144}$ & 2 & $1.99$ & $(1.84,2.16)$ & $2.00$ & $(1.84,2.17)$ & $2.02$ & $(1.85,2.19)$ \\
\bottomrule
\end{tabular*}
\end{table*}

Table \ref{tab1} contains the simulation results for the variance terms $\tau^2_i$ and $\sigma^2_{it}$, with posterior means and 95\% credible intervals calculated over the 2,000 posterior samples. Overall, the posterior means represent the innovation variance terms $\tau^2_i$ very well. For all three variables, $i=1,2,3$, the true values fall within the credible interval. For the measurement error terms $\sigma^2_{it}$, we show the posterior mean estimates and credible intervals for three different time points, $t=1,72,144$, to represent the beginning, middle, and end of the synthetic observations. For all variables, the posterior means are very close to the true parameter value and the posterior credible intervals cover this true value. In addition, we calculated the coverage of the $\sigma^2_{it}$ credible intervals over all time points and variables, and obtained a total coverage of $97.22\%$. This indicates that the $\sigma^2_{it}$ parameters are generally recovered well by the model at all time points.

\begin{figure}[t]
\centering
\includegraphics[width=\linewidth]{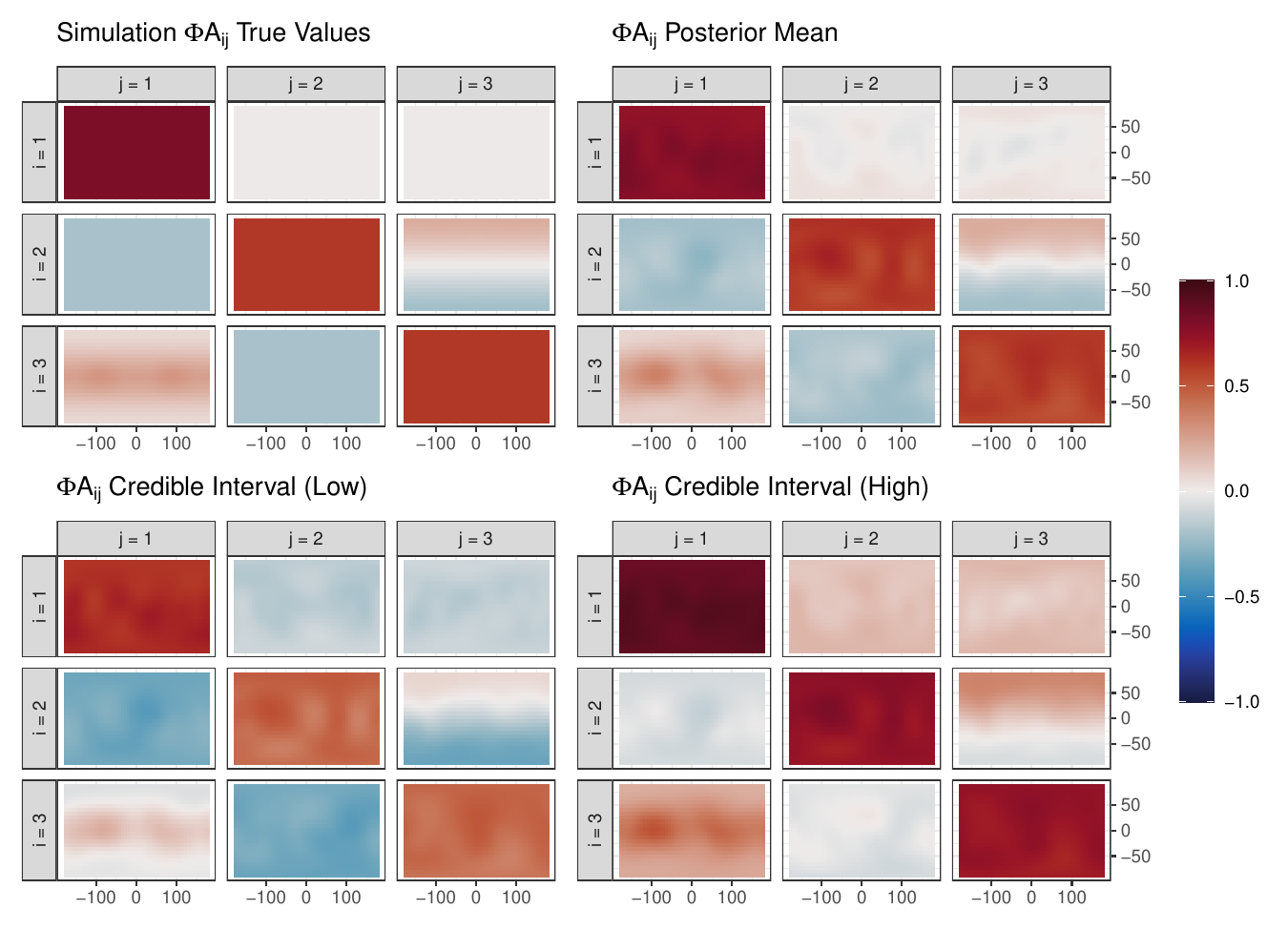}
\caption{
Comparison of true $\bm A$ matrix values and posterior sampling results for a synthetic dataset. In all four panels, $\bm A$ is separated into the blocks $A_{ij}$ and multiplied by the basis $\bm \Phi$ to project the results onto the full spatial domain. Posterior means and credible intervals are calculated separately for each element of $\bm A$ over the 2,000 posterior samples. The lower and upper bounds of the credible interval are based on the 0.025 and 0.975 quantiles of the posterior samples.}
\label{APhi_sim}
\end{figure}

For the transition matrix $\bm A$, Figure \ref{APhi_sim} provides a comparison of the posterior sampling results to the true values. 
The posterior mean recovers the true values of the coefficients in each $ A_{ij}$ block. The only notable difference is some small spatial noise, which is most apparent in blocks such as $A_{23}$, that feature a non-constant spatial pattern. 
In cases where elements of $\bm A$ are large in magnitude, such as the diagonal blocks $A_{ii}, i=1,2,3$, the credible intervals do not contain $0$. 
This is also seen in cases with weaker temporal dependence, such as $A_{21}$ and $A_{32}$. 
For $A_{12}$ and $A_{13}$, the blocks with $0$ temporal dependence, the lower and upper bounds of the credible intervals have a magnitude of about $0.15$.
For the blocks $A_{31}$ and $A_{23}$ that have nontrivial spatial patterns for the true coefficients, the results are similar. For example, the credible intervals for $A_{23}$ contain only positive values near the north pole, negative values near the south pole, and cover 0 near the horizontal line at the equator, matching the spatial trends in the true coefficient values.
Overall, our MCMC sampler does an excellent job at recovering the true values for $\tau^2_i$, $\sigma^2_{it}$, and $\bm A$ given the simulated $\bm Y_t$ data. 

\section{Application}\label{sec:application}

We apply our MV-STDM to study the interaction and evolution of key atmospheric processes following the 1991 Mt. Pinatubo eruption. We focus on three variables from the MERRA-2 Reanalysis datasets: aerosol optical depth (AOD), upwelling longwave radiation at the top of atmosphere (LWR), and 50mb stratospheric temperatures (T50) as described in section \ref{sec:data}. 
It is important to note that we converted each variable to standardized anomalies to remove typical seasonal and spatial trends, also described in section \ref{sec:data}. First, we discuss the various parameter specifications and prior distributions used in our MCMC sampler to model the MERRA-2 anomalies. 
In this section, our focus is twofold: parameter inference, to study the various temporal and multivariate trends present in the data, and prediction over large spatial regions. For the goal of parameter inference, we start by fitting our MV-STDM to the AOD, LWR, and T50 anomalies from 1984-1995. In section \ref{sec:app_multi}, we study the estimated transition matrix to understand the evolution of each process during this time period. To assess prediction performance, we compare our MV-STDM to two univariate spatiotemporal approaches, the univariate version of the MV-STDM for stratosphere temperature that includes the $A$ matrix and a univariate version where the state variables (basis-function coefficients) vary in time thru a random-walk evolution (first-order DLM). 
In section \ref{sec:app_pred}, we present two test set studies: a random hold out set and a spatial block of observations over North America, with the main focus on the latter.

We consider the MERRA-2 AOD, LWR, and T50 anomalies as our multivariate spatiotemporal fields data with $M=3$ variables each observed at a common set of $N=1,152$ spatial locations (on a $24\times 48$ spatial grid) and $T=144$ time points at a monthly frequency. To fit our MV-STDM in (\ref{eq:dlm_full}) to this data, we must specify a few parameters and prior distributions.  
First, we specify the basis, $\bm \Phi$, by generating  $K=162$ RBFs on an icosahedral grid. To parameterize the spatial autoregression matrix $\bm B$, which is a component of the innovation variance $\bm Q$, we use a value of $\kappa=2$.  This parameter choice imposes weak spatial correlation between neighboring RBFs, allowing for spatial flexibility in the state vector $\bm\alpha_t$. 
To complete our model specification, uninformative priors are assigned according to (\ref{sec:mcmc}) with hyperparameters $\bm m_0 = \bm 0_{MK}$, $\bm C_0 = \bm I_{MK}$, $a_\sigma = b_\sigma = 1$, $a_\tau = b_\tau = 1$, and $\lambda = 1/4$. Using our MCMC sampler, we computed two chains with a length of $1,500$ iterations each. Computations for each chain took approximately 51 hours. 
For both chains, satisfactory convergence was diagnosed from trace plots within $250$ iterations. We drop the first $500$ iterations from each chain and combine the remaining iterations to obtain a final set of $2,000$ posterior samples for each parameter. 

\subsection{Multivariate Model Parameter Inference}
\label{sec:app_multi}

\begin{figure}[ht]
\centering
\includegraphics[width=\linewidth]{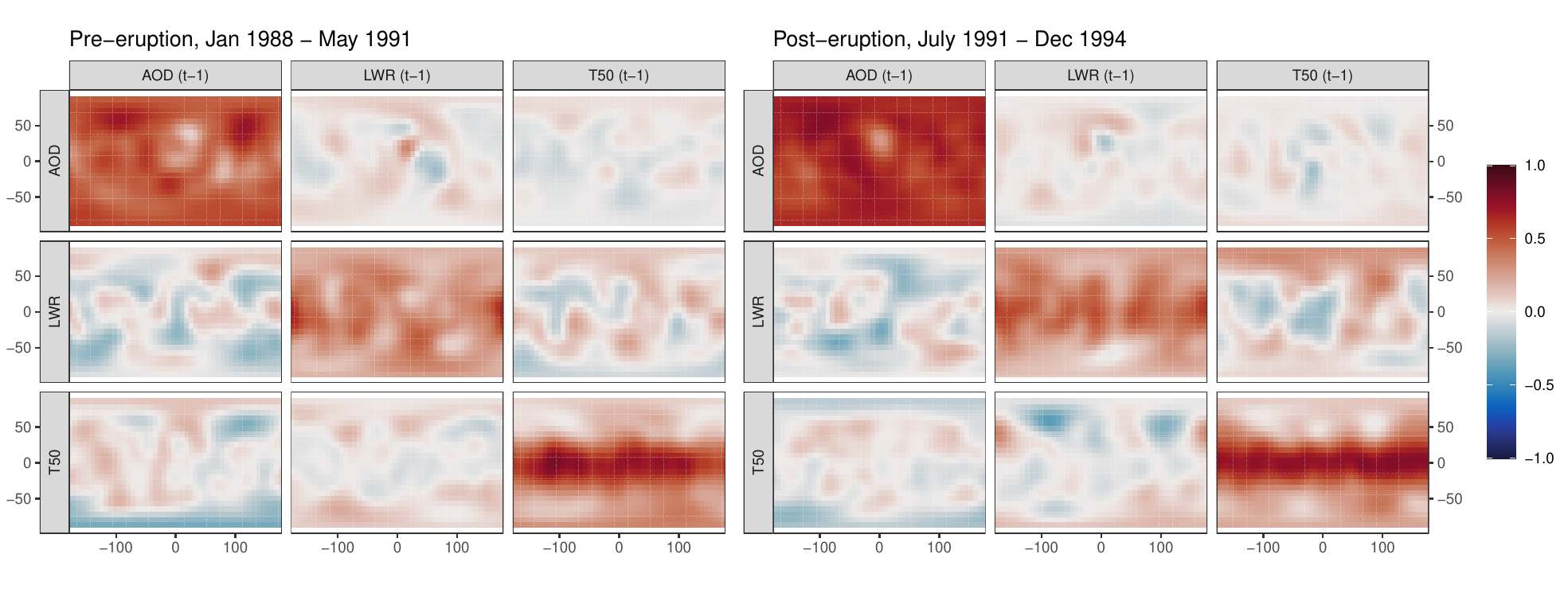}
\caption{
Posterior mean of the transition matrices $\bm A_{pre}$ (left) and $\bm A_{post}$ (right) for the MERRA-2 AOD, LWR, and T50 anomalies.
For interpretation, $\bm A_{pre}$ and $\bm A_{post}$ are separated into blocks $A_{ij}$ and multiplied by the basis $\bm \Phi$ to project the results onto the full spatial domain, represented here by latitude-longitude coordinates on the y-axis and x-axis, respectively. 
Each of the nine blocks represents the relationship between two variables, with the vertical facets denoting the variable at the current month and the horizontal facets denoting the variable at the previous month, denoted as the $t-1$ time step.
Posterior means are calculated separately for each element of $\bm A$ over the 2,000 posterior samples.  
}\label{fig:APhi_application}
\end{figure}

In this part of the study, we focus on parameter inference for elements of the transition matrix, $\bm A$, to provide insights into the spatially-varying temporal and multivariate trends captured by our model. To do this, we train the model using historical AOD, LWR, and T50 anomalies before and after the June 1991 eruption: from January 1988 - May 1991 (41 months pre-eruption) and July 1991 - December (42 months post-eruption). Comparing $\bm A_{pre}$ with $\bm A_{post}$ allows us to understand how auto- and cross-correlations change pre- and post-event, respectively.
As discussed in Section \ref{sec:method_multi}, $\bm A$ is separated into various blocks that represent the temporal dependence for one variable on itself or another variable in the previous month.
For clarity, we denote anomalies from the previous month as the $t-1$ time step.
The posterior mean results for $\bm \Phi \bm A$ pre- and post-eruption are provided in Figure \ref{fig:APhi_application} with each block $A_{ij}$ projected onto the global spatial field using the process detailed in Section \ref{sec:method_multi}. Posterior mean results for coefficients $\bm A_{pre}$ and $\bm A_{post}$ are provided in Appendix \ref{app:amatrix} for completeness.

\begin{figure}[ht]
\centering
\includegraphics[width=\linewidth]{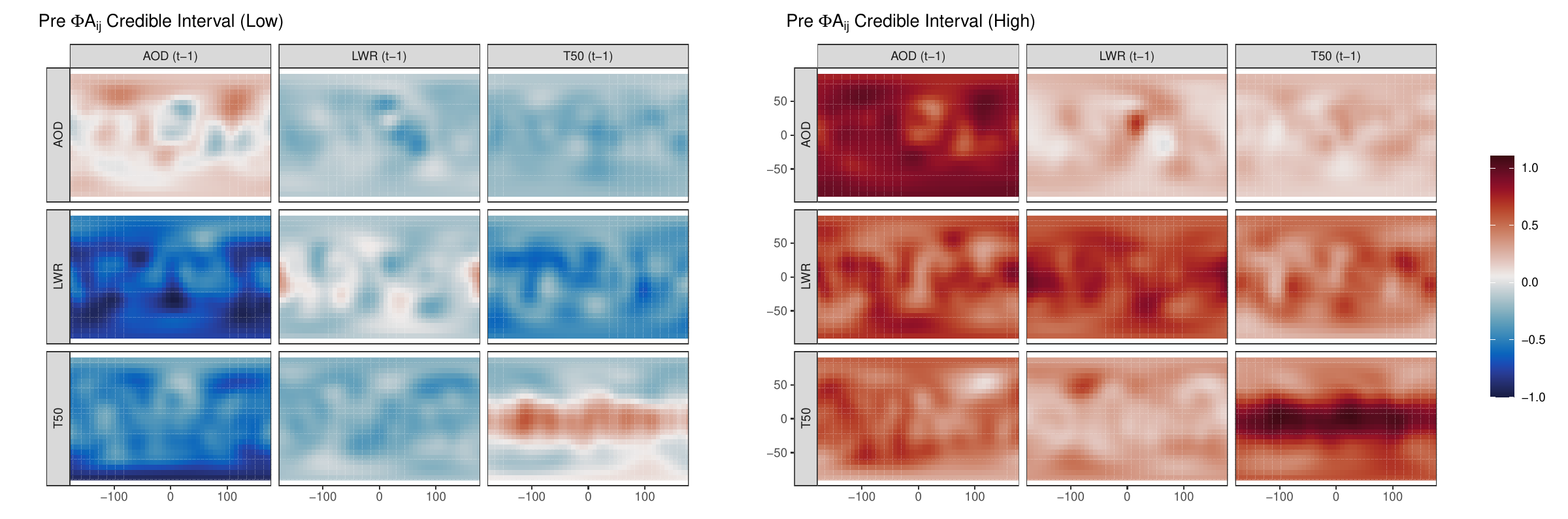}
\includegraphics[width=\linewidth]{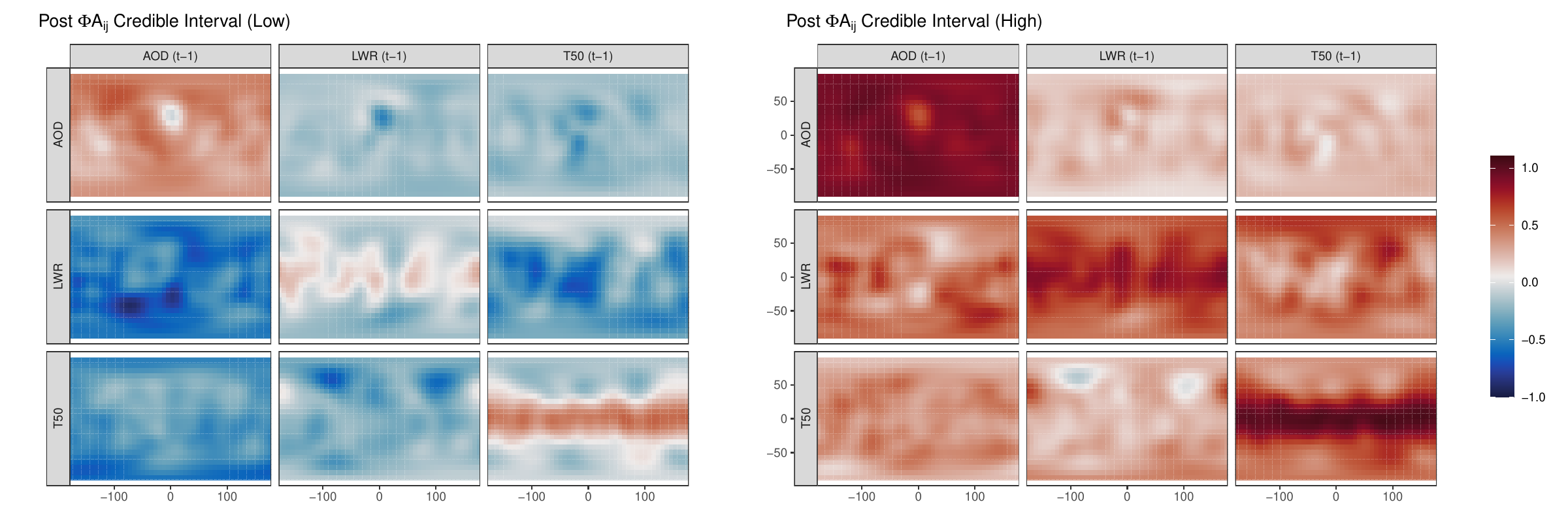}
\caption{
95\% posterior credible intervals for transition matrices, matrices $\bm A_{pre}$ (bottom row) and $\bm A_{post}$ (top row), for the MERRA-2 AOD, LWR, and T50 anomalies.
For interpretation, {$\bm A_{pre}$ and $\bm A_{post}$}} are separated into blocks $A_{ij}$ and multiplied by the basis $\bm \Phi$ to project the results onto the full spatial domain. 
Each of the nine blocks here represents the relationship between one variable and the $t-1$ time step of another variable, and these relationships are allowed to vary over space.
Credible intervals are calculated separately for each element of $\bm A$ over the 2,000 posterior samples, and the lower and upper bounds of the credible interval are respectively based on the 0.025 and 0.975 quantiles of the posterior samples.

\label{fig:APhi_application_ci}
\end{figure}

Consider the three blocks on the diagonals in figure \ref{fig:APhi_application}, which represent the temporal autocorrelations for the state vectors $\bm\alpha^{\text{AOD}}_t$, $\bm\alpha^{\text{LWR}}_t$, and $\bm\alpha^{\text{T50}}_t$.
For all three variables, we see strong positive autocorrelation in almost all regions, especially in the equatorial band between $30^\circ$S and $30^\circ$N pre- and post-eruption.
Across the globe, autocorrelation is strongest for AOD and weaker for LWR and T50.
Both LWR and T50 see the strongest temporal autocorrelations near the equator and mid latitude bands ($30^\circ$N to $60^\circ$N and $30^\circ$S to $60^\circ$S), but these trends taper off, featuring lower but still positive values near the north and south poles. This spatial trend is particularly pronounced for T50.
This is expected as temperature and main drivers of temperature changes are less variable in the equatorial region due to the relatively constant angle between these regions and the sun. More significant changes from season to season, and thus more overall variability, are expected in these variables as one moves further from the equator. 
Autocorrelations, especially for AOD and T50, are stronger (more positive) post-eruption which is also expected since the massive amount of aerosols and their impacts introduced by the eruption lingered for months.
However, there are a few exceptions. First, for AOD, there is one region over northern Africa which exhibits near-zero autocorrelation. This coincides with a neighboring region in central Europe which also exhibits near-zero self dependency for LWR. This pattern is most prominent prior to the eruption.

Next, we consider the off-diagonal blocks of $\bm A$ which feature the lagged cross-dependencies, or relationships between each variable at time step $t$ and the other two variables at time step $t-1$. First, we focus on the top row of Figure \ref{fig:APhi_application}, where the second and third blocks contain the dependence of $\bm\alpha^{\text{AOD}}_t$ on $\bm\alpha^{\text{LWR}}_{t-1}$ and $\bm\alpha^{\text{T50}}_{t-1}$, respectively. Both LWR and T50 at $t-1$ have a near-zero impact on AOD at time $t$ in all regions with one exception (see row 1, columns 2 and 3 in Figure \ref{fig:APhi_application}). 
The regions over northern Africa and central Europe contain compensating negative and positive dependencies, respectively. These match the regions with abnormal autocorrelations for AOD and LWR discussed in the previous paragraph. 
Overall, the near-zero cross-dependencies are not surprising in this case because we expect AOD would have a stronger effect on LWR and T50, but recent studies suggest we should expect some feedback from T050 on LWR and AOD \citep{mielonen2016}. 

We shift our focus to the second row of Figure \ref{fig:APhi_application}, where the first and third blocks feature the dependence of $\bm\alpha^{\text{LWR}}_t$ on $\bm\alpha^{\text{AOD}}_{t-1}$ and $\bm\alpha^{\text{T50}}_{t-1}$. The impact of AOD on LWR pre-eruption is spatially mixed, with negative values in most regions but pockets of smaller positive values near the equator and north pole. The negative values seen in many regions make sense given the expected relationship where stratospheric aerosols absorb outgoing longwave radiation, however the positive values near the equator are unexpected.
The pre-eruption impact of T50 on LWR is also spatially mixed, with positive and negative values dispersed throughout. However, none of these coefficients are particularly strong, so the impact of T50 on LWR seems generally smaller than the impact of AOD on LWR. This makes sense as changes in AOD are expected to sequentially lead to changes in LWR which in turn can impact T50 with some expected feedback between LWR and T50 \citep{mielonen2016}. 
Interestingly, similar spatial patterns can be seen on the second row of the post-eruption results with a notable increase in magnitude of the impact of T50 on LWR, especially in the northen hemisphere where the correlations have become more positive. This indicates that T50 is more postively correlated with LWR post-eruption than pre-eruption. However, the impact of AOD on LWR in the same hemisphere appears to have slightly decreased in magnitude.

The last row of Figure \ref{fig:APhi_application} quantifies the dependence of $\bm\alpha^{\text{T50}}_t$ on $\bm\alpha^{\text{AOD}}_{t-1}$ and $\bm\alpha^{\text{LWR}}_{t-1}$ and is of most interest for this application since stratospheric temperatures are the most long lasting impact from the eruption subsequently leading to significant global surface temperature changes \citep{parker1996}. We notice that AOD has a positive impact on T50 in many regions, especially in the mid-upper latitudes pre-eruption.
Post-eruption results indicate there may be an underlying change in the impact of AOD and LWR on stratospheric temperatures, suggesting a cooling effect due to the eruption in the far northern latitudes but a heating effect in the mid to lower latitudes. This can be seen in the negative impact of AOD and LWR on T50 in the northern latitudes when compared to the positive correlations estimated between the same variables in the pre-eruption results. The isolated heating effect in the lower latitudes is particularly interesting because, as discussed in Section \ref{sec:intro}, there was an observed heating in the stratosphere post-eruption between the equator and 30$^\circ$N \citep{labitzke1994}. The larger negative correlations between LWR$(t-1)$ and T50 in the northern hemisphere may be related to the increased absorption of LWR in that region following the eruption.
This matches the expected trend because these regions saw the greatest levels of stratospheric heating following the Mt. Pinatubo eruption as discussed in Section \ref{sec:intro}. However, the magnitude of these coefficients is small, and in some regions (especially near the south pole), we notice a slight negative dependency of T50 on AOD. 
We see a spatially mixed trend when assessing the dependence of T50 on LWR, with a mix of negative and positive values at the north and south poles, respectively, and near-zero values in the middle latitudes. The magnitude of these coefficients is small across the globe, though the larger negative values in the northern hemisphere may be related to the increased absorption of LWR in that region following the eruption. 


Overall, the posterior mean estimates of the temporal autocorrelations seem strong, most notably at the equator and lower latitudes, but the lagged cross-dependencies exhibit a higher degree of uncertainty. This is reflected in the 95\% credible intervals for $\bm A$ shown in Figure \ref{fig:APhi_application_ci}.  
For both the pre- and post-eruption results, the credible intervals for the autocorrelations (diagonal blocks) do not contain zero in the regions with strong positive values, but do contain zero in the regions with smaller values. This shows evidence of significant spatially-varying trends in the evolution of each process. 
For the lagged cross-dependencies, represented by the off-diagonal blocks, the values are relatively close to 0 and the credible intervals do contain 0 in most cases. There is also more uncertainty in these parameter estimates represented by wider credible intervals. A notable exception are the blue regions in the northern hemisphere over North America and Russia in the post-eruption credible intervals representing the impact of LWR on T50.
These results suggest that the multivariate trends in the MERRA-2 anomalies may be too small in magnitude to recover with our MV-STDM, or that our model may be overspecified. If the latter is the case, the parameter estimates for $\bm A$ may account for too much of the natural variability in the data, thus obscuring the underlying trends.
The simulation results in Section \ref{sec:simulation} indicate that for this data dimension and a similar model specification, the credible intervals produced by our MCMC sampler can be somewhat wide even in the case where the trends in $\bm A$ are clear. So, parameters in $\bm A$ which are small in magnitude could still represent underlying trends in the anomalies.
Given that some of the posterior means in the off-diagonal blocks of Figure \ref{fig:APhi_application} do feature clear spatial trends despite the overall low values, the multivariate model may still contain useful information on the interaction between atmospheric processes. 

\subsection{Assessment of predictive performance}
\label{sec:app_pred}

While parameter inference to characterize the evolution of atmospheric processes and understand their interactions is the primary focus for our application, we also want to assess our model's capability in predicting downstream impacts of the eruption in localized regions. We assessed prediction performance across three models and two distinctive spatial hold out sets: a random $10\%$ of locations and a spatial block over North America. 
Predictions for the random hold-out set showed no clear performance advantage for any of the three models (see Appendix for results). Although uninteresting, this was a necessary sanity check and expected outcome because of the large amount of spatiotemporal information that random observations have to draw from. This is especially the case for continuous geospatial data, such as the MERRA-2 data as opposed to sparse ground measurements, for example. We might expect the performance of the univariate cases to decline as the test/training data ratio increases. On the other hand, the spatial block hold out set demonstrates when spatial and temporal information is insufficient and multivariate correlations can significantly add value. We will focus on the results of the spatial block test set for the remainder of this section. 

\begin{figure}[hb]
\centering
\includegraphics[width=0.85\linewidth]{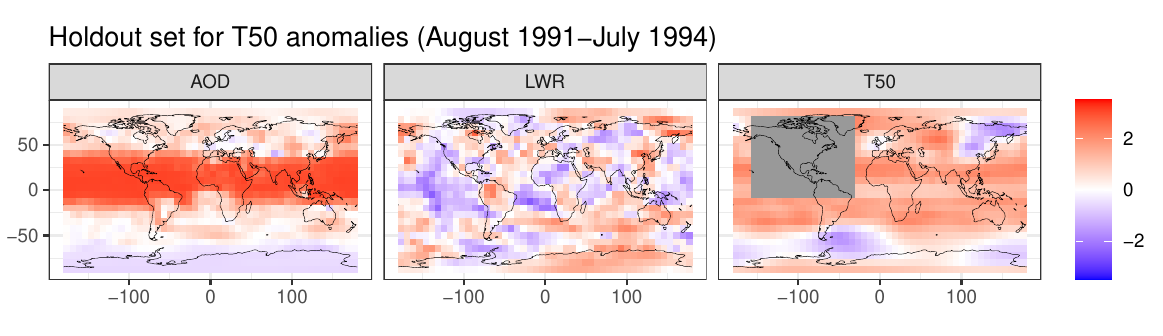}
\caption{
Map of the holdout set of observations for stratospheric temperature (T50). The gray region represents the region over which observations are excluded for the August 1991-July 1994 time period when fitting the model. These observations are later used to assess the model's predictive performance. The anomalies shown in the map are for August 1991.
\label{fig:valid_scene}
}
\end{figure}

We test our model's performance in predicting each variable over North America for a three-year period following the eruption (August 1991-July 1994) using the other two variables. For stratospheric temperature, this region is shown in Figure \ref{fig:valid_scene}, and covers the area bounded by $155^\circ$W to $35^\circ$W longitude and $5^\circ$S to $80^\circ$N latitude.
This holdout set represents a large missing block of observations in both space and time for a single variable and thus highlights the benefit of our multivariate model.
This swath of missing data also mimics patterns of missingness often present in truly observational satellite data such as that collected near-instantaneously by the atmospheric sensor Moderate Resolution Imaging Spectroradiometer \citep[MODIS,][]{justice2002}
aboard polar orbiting Aqua and Terra satellites.
Note that such data did not become available until after 1995 and thus did not measure impacts from the Mt Pinatubo eruption. However, the interpolation of missing observations is one area in which our model may be useful for understanding the impact of future eruption events. In particular, good predictability can provide early insights into potential near-term impacts after a recently observed event. 

We focus on three models for comparison. First, we consider the full MV-STDM discussed in Section \ref{sec:app_multi} with all three variables (AOD, LWR, T50). Then, to evaluate our model's utility as a (univariate) spatiotemporal dynamic model, we fit a second model, considering only the AOD, LWR or T50 anomalies. While this model only considers temporal dependencies for a single variable, it still offers novel contributions compared to previous approaches due to the spatially-varying structure of the transition matrix. Lastly, we consider a simplified univariate model where we constrain $\bm A = \bm I_K$, again for a single variable only. We denote these three models as ``Multivariate'', ``Univariate'', and ``Univariate-Random Walk'', respectively. We call the last model ``Univariate-Random Walk'' or ``Univariate-RW'' because it represents a first-order DLM structure \citep{petris2009} where the transition equation reduces to $\bm \alpha_t = \bm\alpha_{t-1} + \bm \eta_t$. 
This simplified model creates a nonstationary (and thus flexible) temporal structure and eliminates the need to estimate additional model parameters, and is thus popular for scalable spatiotemporal models \citep[i.e.][which estimate transition coefficients for the regression parameters, but not the spatial random effect]{finley2012bayesian, datta2016}. 
For all three models, we follow the specifications discussed at the beginning of Section \ref{sec:application} with the exception of $\bm m_0$ and $\bm C_0$, the prior hyperparameters for $\bm A$. For the Multivariate model, these hyperparameters remain the same, but for the Univariate model, we must reduce the dimension of these hyperparameters to match the size of the univariate state vector, so we specify $\bm m_0=\bm 0_K$ and $\bm C_0=\bm I_K$. For the Univariate-RW model, $\bm A$ is not estimated so the hyperparameters $\bm m_0$ and $\bm C_0$ are not needed. Two chains of 1,500 MCMC iterations are computed for each submodel, and the first 500 iterations are dropped as burn-in to provide a final set of 2,000 posterior samples. In each case, we exclude the data in the holdout set when training the model. Interpolation for the missing observations occurs naturally using the spatial basis function representation in the observation equation.

\begin{figure}[ht]
\centering
\includegraphics[width=0.8\linewidth]{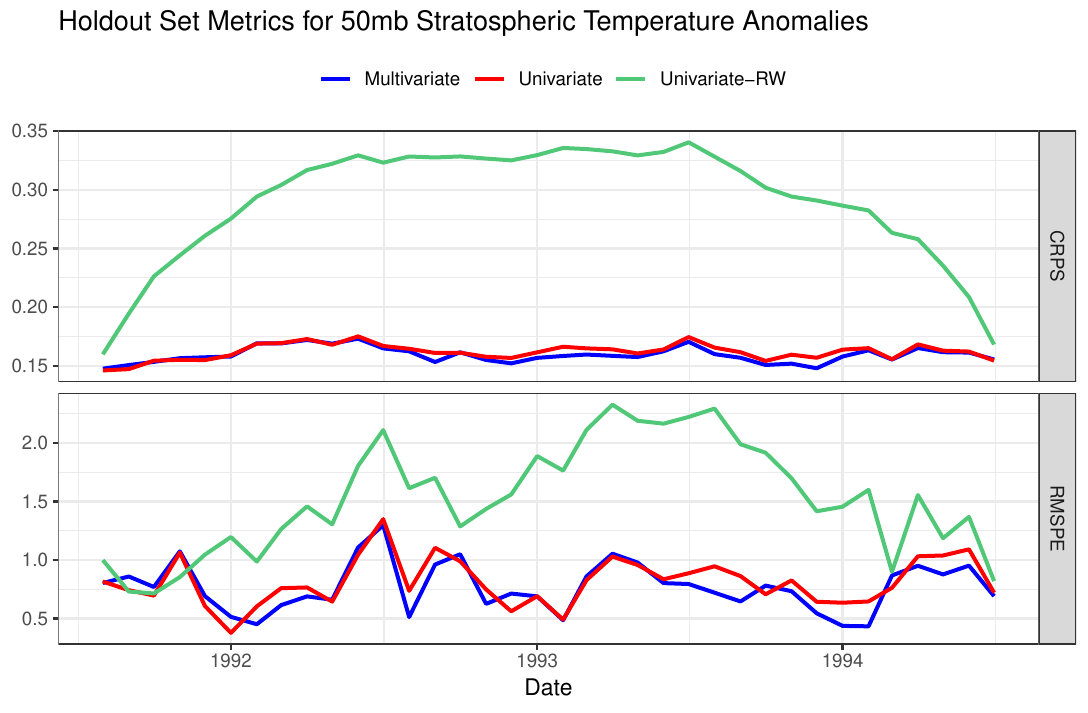}    
\caption{Predictive performance for the three submodels on the stratospheric temperature holdout set shown in Figure \ref{fig:valid_scene}. Results are presented for each month from August 1991 through July 1994. The top panel contains the continuous ranked probability score, or CRPS, while the bottom panel contains the root mean squared prediction error, or RMSPE, for each month. Different colors are used to represent each submodel. Similar figures for predicting AOD and LWR are shown in the appendix.} 
    \label{fig:metrics}
\end{figure}

\begin{table}[t]
\centering
\begin{tabular}{|l|cc|cc|cc|}
    \hline
    &  \multicolumn{2}{c|}{\textbf{AOD}} & \multicolumn{2}{c|}{\textbf{LWR}} & \multicolumn{2}{c|}{\textbf{T50}} \\
    \textbf{Model} & CRPS & RMSPE & CRPS & RMSPE & CRPS & RMSPE  \\
    \hline
    Multivariate  & 0.118 & 1.438 & 0.245 & 1.583 & 0.159 & 0.796 \\
    Univariate    & 0.128 & 1.398 & 0.284 & 1.540 & 0.162 & 0.836\\
    Univariate-RW & 0.197 & 1.584 & 0.466 & 1.901 & 0.291 & 1.596\\
        \hline
\end{tabular}
\caption{Metric average over the entire time period for each holdout variable.}\label{tab:results}
\end{table}

For each model, we use two metrics to assess predictive performance: root mean squared prediction error (RMPSE) and continuous rank probability score \citep[CRPS,][]{gneiting2007}. Both metrics are computed based on the posterior sampling results for all observations in the holdout set. RMSPE provides a direct assessment of accuracy based on the mean of the posterior predictive distribution, while CRPS provides a more comprehensive evaluation based on the entire posterior predictive distribution.
For each month in the hold out set from August 1991-July 1991, we compute our predictive metrics using all locations in the holdout region to assess the performance of each submodel over time. 
Each RMSPE calculation is performed by taking the square root of the average squared prediction error over the entire holdout set, whereas CRPS is first calculated at each location using the empirical cumulative distribution-based approximation \citep[discussed in][]{Kruger2021} and later averaged over the holdout set. 
To provide a single number summary, we perform similar calculations of each metric over the entire time period of August 1991-July 1991.

Table \ref{tab:results} summarizes the time-averaged metrics for each variable held out.
When averaging over the entire time period, our Multivariate model provides the best predictive performance across both metrics in most cases with the Univariate model a close second. For AOD and LWR, CRPS indicates that the Multivariate model performs best but RMSPE indicates the Univariate model performs as well or slightly better. This result is interesting and makes sense because these variables, AOD especially, exhibit strong autocorrelation across all locations indicating that information from other variables contributes little additional value.
The simplified Univariate-RW model exhibits poorer predictive performance with higher values for both metrics across all three variables. 
The monthly results for predicting the stratospheric temperature hold out block are shown in Figure \ref{fig:metrics}. 
When looking at predictive performance over time for temperature, the Multivariate and Univariate models appear to share similar performance each month, especially for CRPS, though the Multivariate model tends to exhibit superior performance most months by a small margin.
Whereas, the Univariate-RW model achieves its best performance close to the edges of the holdout time period, where it shares similar RMSPE values and slightly higher CRPS values than the other models, with performance degrading for both metrics in the middle time periods. 
See section \ref{app:pred_othervars} in the appendix for monthly predictions results for AOD and LWR which show a similar trend for CRPS.
Our previous results in Section \ref{sec:app_multi} show significant evidence that the $\bm A$ matrix coefficients representing the autocorrelation of the AOD, LWR and T50 anomalies, are less than 1, suggesting a significant difference in temporal dynamics compared to the Univariate-RW model. 
These coefficients also vary over space, unlike the Univariate-RW model, allowing the model to adapt to changes in the evolution of the T50 anomalies in different regions.
The Univariate-RW model likely struggles with our holdout scenario because it does not share these features. 
In most cases, our results in Section \ref{sec:app_multi} suggest that the Multivariate model struggled to capture strong cross-correlations between variables over time, so it is overall not surprising to see that our univariate model performs almost as well as our Multivariate model. However, the increased predictive performance of the Multivariate model, though slight, does provide evidence that our method can successfully capture and leverage information about the joint evolution of atmospheric processes. 

\section{Discussion}\label{sec:discussion}

Understanding interactions between atmospheric processes is essential for characterizing and anticipating the risks and impacts associated with climate intervention efforts such as SAI. To that end, we built a multivariate space-time dynamic model (MV-STDM) which offers a novel and interpretable parameterization to capture the joint evolution of spatiotemporal processes inherent to our atmospheric data around the 1991 Mt Pinatubo volcanic eruption. 
We developed a customized MCMC sampler to estimate our model's parameters for multivariate spatiotemporal fields data and established it's effectiveness on a simulated dataset. 
Finally, we applied our model to characterize the multivariate evolution of aerosol optical depth, longwave radiation, and stratospheric temperatures before and after the 1991 Mt. Pinatubo eruption.
Through parameter inference, we discovered significant spatial trends in the transition matrix and gained insight on the interaction between processes that corroborated known science. We also discovered notable changes in the transition matrix before and after the event, highlighting the value of our approach to make meaningful inference on how an SAI event could change the near-term or long-term climate.
Additionally, our results demonstrated improved predictive performance when using our proposed MV-STDM when compared to simpler model specifications. 
We found this performance especially enhanced when the test set was a large spatial block, emphasizing the importance of information from other variables when spatial and temporal correlation alone is insufficient. We found that the multivariate information was more critical to prediction over large spatial regions than in the case of random spatial hold out sets.

The analysis in Section \ref{sec:app_multi} shows that our MV-STDM recovers known features of the temporal evolution of AOD, LWR, and T50, such as the strong autocorrelation focused near the equator. However, these results do not always match our expectations and can exhibit a large degree of uncertainty. For example, the strong interaction between AOD and longwave radiation over northern Africa is likely an artifact of overfitting to the anomalies in that region. Because AOD in this region is predominately driven by Saharan dust events, which feature abnormal temporal characteristics in terms of their frequency and duration \citep{goudie2001}, our time period of 1984-1995 may not be long enough to accurately calculate the climatologies (and therefore anomalies) in this region to account for these events. The addition of geographically relevant covariates could also be useful here such as the inclusion of land or vegetation cover.
Similarly, it is impossible to isolate a single event from observational data. Advanced methods that could better capture the multiple sources of natural variability such as the El Ni{\~n}o Southern Oscillation, could further draw out the signal from the noise \citep[e.g][]{thompson2009}. This could be addressed through an alternate basis function representation or additional data processing steps. 

Our model's linear, first order evolution equation was chosen for interpretability and scalability, but additional time steps or a nonlinear formulation could aid in capturing more complex spatiotemporal dynamics. However, the addition of other time steps in the evolution equation would lead into a model formulation incompatible with the traditional FFBS algorithm. 
Another option would be to add a ``coregionalization'' structure to the innovation variance similar to \cite{gelfand2005}. This would allow for multivariate dependencies in the innovation of the basis coefficients, capturing the contemporaneous, rather than time-lagged, relationship between variables. This would introduce greater model complexity potentially hindering interpretability, but could provide useful insight into the nature of the interactions between observed atmospheric processes. 

Compared to  Lattice Kriging \citep{nychka2015}, our analysis uses a relatively small number of basis functions, and increasing the resolution further may be impractical given our MCMC sampling approach. Improving computational speed is thus a key focus for future work. While Lattice Kriging employs sparse Cholesky decompositions to estimate basis coefficients, our FFBS algorithm for estimating the $\bm \alpha_t$ coefficients faces challenges due to the need for solving multiple linear systems sequentially. Unlike Lattice Kriging, we do not estimate the $\kappa$ parameter to maintain computational tractability over multiple variables and MCMC iterations. 
Other spatial representations or approximations may help to enhance the scalability of our model. The spatiotemporal FFBS approach of \cite{jurek2023} is one potential avenue, although its suitability for RBF models is unexplored. Replacing the RBFs with spectral representations or empirical orthogonal functions (EOFs) could allow for more fine-scale spatial information without increasing the basis dimension, but might hinder interpretation of the transition matrix. 
Another alternative to our MV-STDM would be to consider the development of a multivariate model space time model via a low-rank representation as in \cite{nguyen2014spatio}. Similar to \citep{nychka2015}, this other approach is capable of handling very large data sets and performing prediction of spatiotemporal fields based on data sources of different resolutions.
Future research will explore these options and their impact on model performance.

Finally, we see many possibilities for future analyses and applications with our MV-STDM.  We could examine changes in the relationships between AOD, LWR, and T50 over time by estimating model parameters over different time windows. Applying the MV-STDM to earth system model (ESM) outputs would allow for comparisons of multivariate interactions between simulations and observations. This approach could also be applied to many runs from the same ESM to assess the impact of internal variability on these atmospheric interactions. Similar analyses could be performed to examine other impacts of the Mt Pinatubo or more broadly any climate interactions of interest.

\section*{Acknowledgments}

\paragraph{Funding Statement}
This paper describes objective technical results and analysis. Any subjective views or opinions that might be expressed in the paper do not necessarily represent the views of the U.S. Department of Energy or the United States Government. This work was supported by the Laboratory Directed Research and Development program at Sandia National Laboratories, a multi-mission laboratory managed and operated by National Technology and Engineering Solutions of Sandia, LLC, a wholly owned subsidiary of Honeywell International, Inc., for the U.S. Department of Energy's National Nuclear Security Administration under contract DE-NA0003525. 

\newpage
\appendix

\section{MCMC Sampling Algorithm} \label{app:implementation}

Following the model equations in (\ref{eq:dlm_full}), we initialize the parameters to some starting values such as
\begin{equation} 
\sigma^2_{it} = 1, \quad 
\tau^2_i = 1, \quad 
\bm A = \bm I_{MK} \quad
\bm \alpha_t = \bm Z_t,
\end{equation}
where $\bm Z_t,$ $t=1,\ldots,T$ is a $KM\times 1$ matrix with each element drawn independently from a standard normal distribution. We then proceed to sample from the following conditional posterior distributions for each MCMC iteration:
\begin{enumerate}

    \item For $i=1,\ldots,M$, $\tau^2_i \sim IG\left(a^*_i,b^*_i\right)$, where
    \begin{align*}
        a^*_i &= a_\tau + \frac{KT}{2} \\
        b^*_i &= b_\tau + \frac{1}{2}\left(\bm \eta_t^{(i)'}\bm B'\bm B\bm \eta_t^{(i)}\right)
    \end{align*}
    
    \item For $t=1,\ldots,T$ and  $i=1,\ldots,M$, $\sigma^2_{it}\sim IG\left(a^*_{it},b^*_{it}\right)$, where
    \begin{align*}
        a^*_{it} &= a_\sigma + \frac{N}{2} \\
        b^*_{it} &= b_\sigma + \frac{1}{2}\left(\bm Y^{(i)}_t-\bm\Phi\bm\alpha^{(i)}_t\right)'\left(Y^{(i)}_t-\bm\Phi\bm\alpha^{(i)}_t\right)
    \end{align*}

    \item Reformulate the parameters of the  measurement equation as in (\ref{eq:mvlr}), then sample $\text{vec}(\bm{\tilde{A}})\sim N(\bm m_A,\bm C_A)$, where
   
       \begin{align*}
        \bm C_A^{-1} &= \begin{bmatrix}
                          \bm{X}^\alpha_{T-1} \\
                          \vdots \\
                          \bm{X}^\alpha_{0} \\
                        \end{bmatrix}'
                        \left(\bm I_T \otimes \bm{Q}^{-1}\right)
                        \begin{bmatrix}
                          \bm{X}^\alpha_{T-1} \\
                          \vdots \\
                          \bm{X}^\alpha_{0} \\
                        \end{bmatrix} + \bm V_0 \\
        \bm m_A &= \bm C_A\left(
                        \begin{bmatrix}
                          \bm{X}^\alpha_{T-1} \\
                          \vdots \\
                          \bm{X}^\alpha_{0} \\
                        \end{bmatrix}'
                         \left(\bm I_T \otimes \bm{Q}^{-1}\right)
                         \begin{bmatrix}
                           \bm\alpha_T \\
                           \vdots \\
                           \bm\alpha_1
                         \end{bmatrix}
                         +\bm V_0 \bm\mu_0\right),
    \end{align*}

where $\bm\mu_0$ and $\bm V_0$ are the prior mean and variance for $\text{vec}(\bm{\tilde{A}})$. Because the innovation variance matrix $\bm Q$ is computed based on the sparse spatial autoregression matrix $\bm B$, we were able to efficiently compute Cholesky factors of $\bm Q^{-1}$. By further leveraging the block diagonal structure of (\bm $I_T \otimes \bm Q^{-1}$), we can compute $\bm C_{A}^{-1}$ as shown above without a costly inversion.

\item Jointly sample $\bm\alpha_0,\ldots ,\bm\alpha_T$ using the FFBS algorithm from \cite{carter1994}, \cite{fruwirth1994}, described in \cite{petris2009} and \cite{West-Harrison97}. This algorithm handles missing observations in $Y_t$, and interpolation for these observations occurs naturally in the measurement equation. Our code implementation for this step is based on the code in the `dlm' R package \citep{petris2010}, but includes modifications for the specific structure of our model.
First, their algorithm requires computing the Singular Value Decomposition (SVD) of each $\bm V_t$ matrix, which is slow for our large spatial model because dense matrices are expected. Because each $\bm V_t$ matrix is diagonal in our model, we directly compute the SVD using an analytical form. Similarly, the `dlm' implementation of FFBS expects dense matrices for other model parameters such as $\bm V_t$ and the design/transition matrices (in our case $\bm \Phi$ and $\bm A$). In our spatial model, these matrices are large in dimension but sparse in structure. So, we opted to use the Rcpp `Armadillo' package, which features the \verb|SpMat| class to handle sparse matrices. This saves memory usage compared to ordinary C++ matrices and also saves computation time because many steps in the FFBS algorithm require taking matrix products of model parameters such as $\bm V_t$, $\bm \Phi_M$, and $\bm A$ \citep{petris2010}.
\end{enumerate}

\section{Transition Matrix Specification for Synthetic Dataset}
\label{app:simulation}

Section \ref{sec:sim_data} discusses the parameter choices for our synthetic data example. Here, we discuss in further detail the parameter specification for $\bm A$, which is depicted in Figure \ref{fig:A_true_sim}.
For this parameter, we specify different types of relationships for each block $A_{ij}$, each of which determines the temporal dependence of variable $i$ on variable $j$. For the autocorrelations, which represent temporal dependencies of each variable on themselves, we specify strong positive values as follows
\begin{equation*}
     A_{11} = 0.8*\bm I_K, \qquad A_{22} = A_{33} = 0.6*\bm I_K.
\end{equation*}
In each case we use a scalar multiple of the identity matrix, so the coefficients are the same for every RBF and thus the pattern of autocorrelation is spatially constant.
For the lagged cross correlations, or temporal dependencies of each variable on the other variables, we consider a variety of coefficients and spatial patterns across the different blocks. To test how our model handles spatially constant negative and zero coefficients, we specify
\begin{equation*}
    A_{12} = A_{13} = 0*\bm I_K, \qquad A_{21}= A_{32} = -0.2*\bm I_K.
\end{equation*}
Then, to test spatially-varying dependencies, we introduce a helper function, $\text{lat}(k)$, which provides the latitude coordinate in degrees of the center point of location $k =1,...,K$ in the basis grid. For the remaining blocks, we first specify
\begin{equation*}
    A_{31} = 0.4*\text{diag}({\tilde{A}}_{31}), \qquad {\tilde{A}}_{31}[k] = 1-\sqrt{|\text{lat}(k)/90|},
\end{equation*}
where ${\tilde{A}}_{31}[k]$ denotes the $k$-th element of the $K\times 1$ vector ${\tilde{A}}_{31}$. The resulting spatial trend has a value of $0.4$ at the equator, but decays to $0$ at the north and south poles. Finally, to test a mix of positive and negative dependencies over space, we specify 
\begin{equation*}
    A_{23} = 0.3*\text{diag}({\tilde{A}}_{23}), \qquad {\tilde{A}}_{23}[k] = \text{lat}(k)/90,
\end{equation*}
where ${\tilde{A}}_{23}[k]$ denotes the $k$-th element of the $K\times 1$ vector ${\tilde{A}}_{23}$.
This provides a linear trend starting at $0.3$ at the north pole and decreasing to $-0.3$ at the south pole.

\section{Autoregressive coefficients pre- and post-eruption}
\label{app:amatrix}

In this section, we include the posterior estimates of the autoregressive coefficients $A_{ij}$ for fitted MV-STDM models pre- and post-eruption. Each dot represents the center of a radial basis function. One interesting pattern we noticed is the negative coefficient (blue dot) around 0$^\circ$ longitude and 25$^\circ$N latitude. This is also seen in the pattern of projected coefficients $\bm \Phi\bm A$ in Figure \ref{fig:APhi_application}.

\begin{figure}[ht]
\centering
\includegraphics[width=\linewidth]{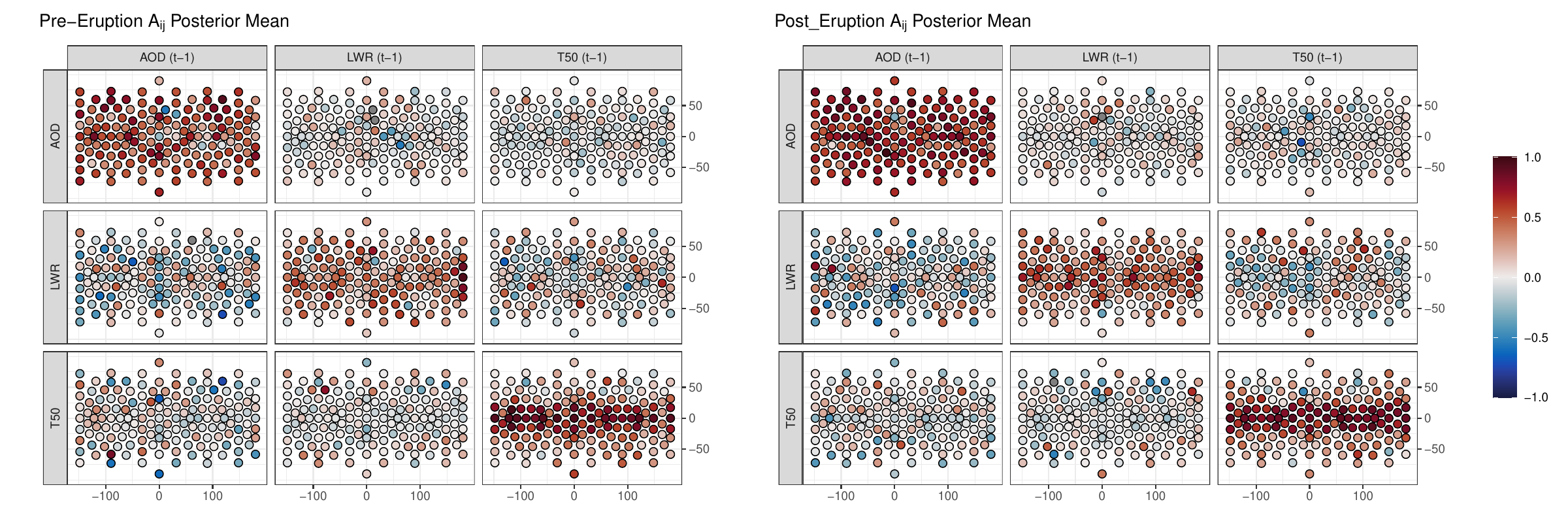}
\caption{Posterior means of spatially-varying autoregressive coefficients $A$ for fitted model pre- and post-eruption. Locations represent the centers of the radial basis functions.} 
    \label{fig:a_coeffs}
\end{figure}

\newpage
\section{Spatial block holdout results for AOD and LWR}
\label{app:pred_othervars}

\begin{figure}[b]
\centering
\includegraphics[width=0.8\linewidth]{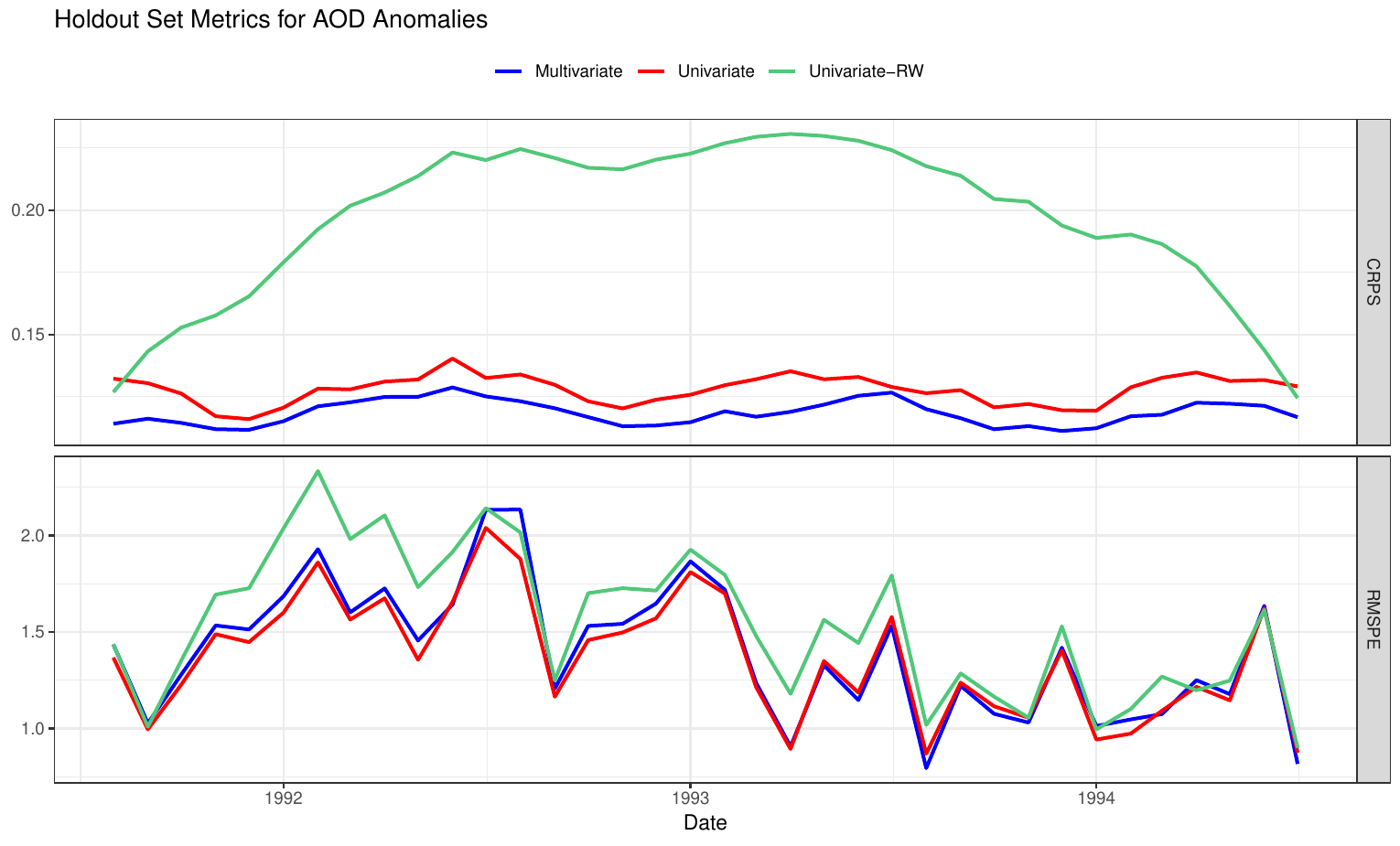}
\includegraphics[width=0.8\linewidth]{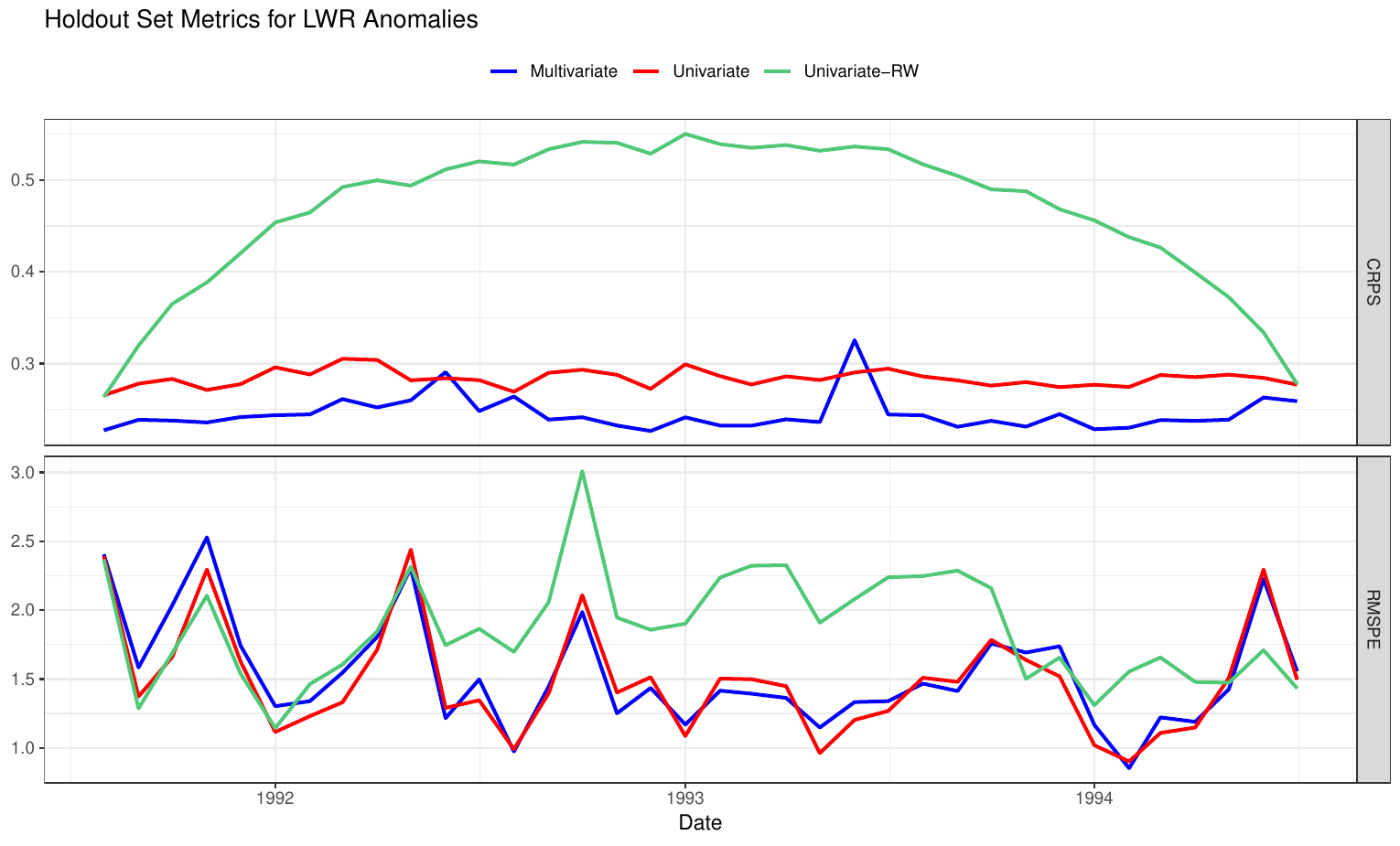}
\caption{Predictive performance for the three submodels on the AOD (top) and LWR (bottom) holdout set using the same missing spatial block shown in Figure \ref{fig:valid_scene}. Results are presented for each month from August 1991 through July 1994. The top panel contains the continuous ranked probability score, or CRPS, while the bottom panel contains the root mean squared prediction error, or RMSPE, for each month. Different colors are used to represent each submodel.} 
\end{figure}

\newpage
\section{Random Holdout Scenario}
\label{app:random_holdout}

\begin{figure}[h]
\centering
        \begin{subfigure}{0.675\linewidth}
        \centering
        \includegraphics[width=\linewidth]{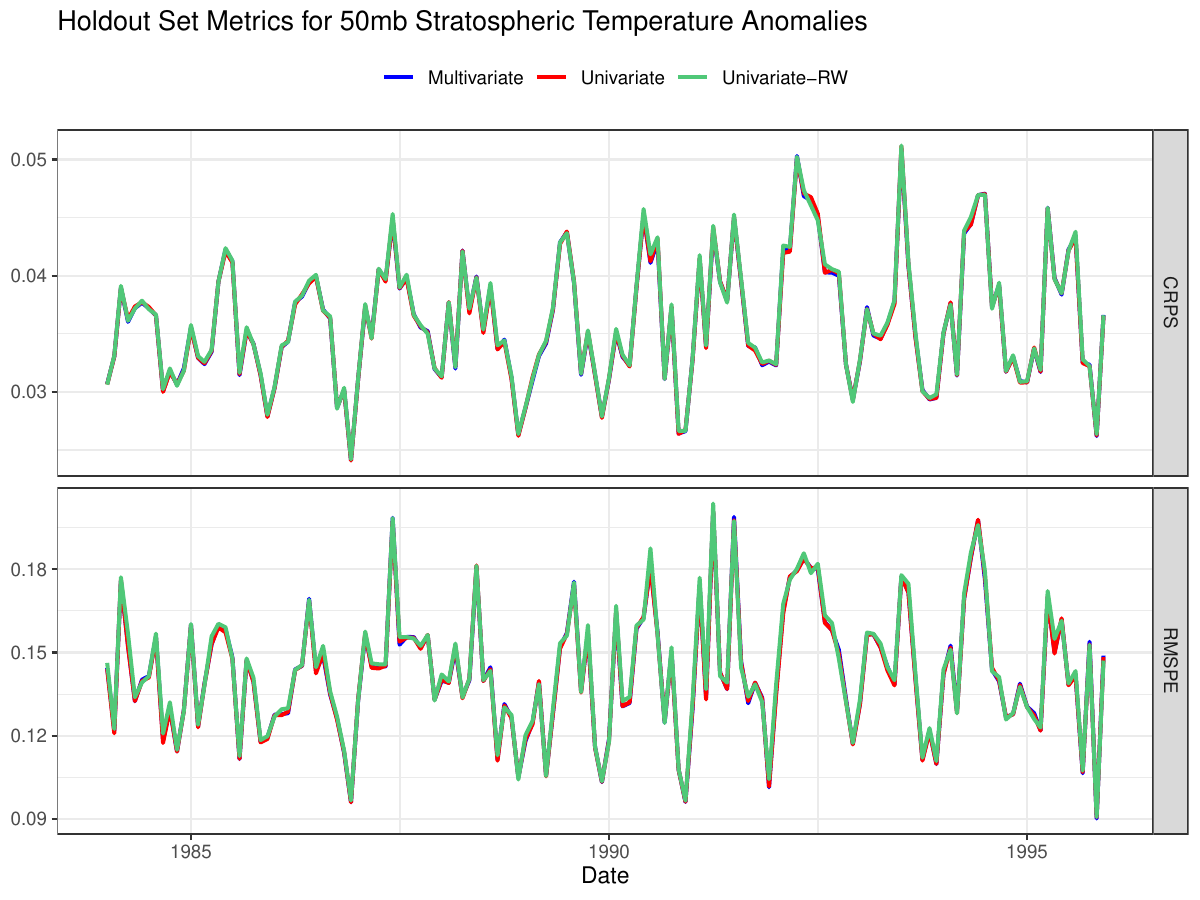}
      \vspace{-2mm}
        \end{subfigure}
        \begin{subfigure}{0.55\linewidth}
        \centering
        \begin{tabular}{l|cc}
        \hline
        \textbf{Model} & \textbf{CRPS} (all months) & \textbf{RMSPE} (all months) \\
        \hline
        Multivariate  & $0.036098$ & $0.144654$ \\
        Univariate    & $0.036103$ & $0.144507$ \\
        Univariate-RW & $0.036213$ & $0.145414$ \\
        \hline
        \end{tabular}
        \end{subfigure}
        
    \caption{Predictive performance for the three submodels on a random set of holdout locations for stratospheric temperature. Results are presented for each month from January 1984 through December 1995. The top panel contains the continuous ranked probability score, or CRPS, while the bottom panel contains the root mean squared prediction error, or RMSPE, for each month. Different colors are used to represent each submodel. The table provides averages for each metric over the entire time period.}
    \label{fig:metrics_s2}
\end{figure}

We test our model's predictive performance for T50 anomalies on a random holdout set of locations over the global domain. This set contains 115 locations, or approximately 10\% of the locations in the full data. At each location, data is held out each month from 1984-1995.
See Section \ref{sec:app_pred} for a description of the models and metrics considered here, and see Figure \ref{fig:metrics_s2} for the results. 

\bibliographystyle{unsrtnat}
\newpage
\bibliography{ref}  

\end{document}